\documentclass[11pt, a4paper]{article}
\pdfoutput=1
\usepackage{jheppub}
\usepackage{amsmath}
\usepackage{graphicx}% Include figure files
\usepackage[config=altsf]{subfig}
\usepackage{feynmf}
\usepackage{dsfont}
\usepackage{array}
\usepackage{slashed}
\usepackage{slashed}
\usepackage{afterpage}
\usepackage{comment}
\usepackage{appendix}
\usepackage{multirow}
\usepackage{dcolumn}
\usepackage{slashed}
\usepackage{soul}
\usepackage{cleveref}
\crefname{figure}{Fig.}{Figs.}

\usepackage{booktabs}    
\usepackage{placeins}
\usepackage{bbm}
\usepackage{amsfonts}
\usepackage{amsmath,amssymb}
\usepackage{mathrsfs}
\usepackage{epsfig}
\usepackage{graphicx}
\usepackage{wrapfig}   

\usepackage{url}
\usepackage{array}
\usepackage{hyperref}
\usepackage{float}
\usepackage{pstricks}
\usepackage{color}
\usepackage{multirow}
\usepackage{multicol}
\usepackage{lipsum}
\usepackage{enumitem}
\usepackage{ctable}
\usepackage{pgfkeys, pgfsys, pgfcalendar}

\vspace*{3cm}
\title{\bf NLO QCD effects on angular observables in $e^-p \to e^-(\nu_e)Hj$ in presence of non-standard $HVV$ couplings}
\author[a]{Biswajit Das,} %\thanks{agrawal@iopb.res.in}}
\author[b]{Pramod Sharma}
\author[b]{and Ambresh Shivaji}

\affiliation[a]{The Institute of Mathematical Sciences,
IV Cross Road, Taramani, Chennai 600113, India}
\affiliation[b]{Department of Physical Sciences, Indian Institute of Science Education and Research Mohali,
Knowledge city, Sector 81, Manauli, PO, Sahibzada Ajit Singh Nagar, Punjab, India, 140306}
\emailAdd{biswajitd@imsc.res.in}
\emailAdd{ph19010@iisermohali.ac.in}
\emailAdd{ashivaji@iisermohali.ac.in}

\abstract{
The single Higgs production in neutral-current (NC) and charged-current (CC) processes at an electron-proton ($ep$) collider is a useful channel to probe new physics effects in the Higgs coupling to vector boson ($HVV$). In this context, observables sensitive to non-standard couplings previously studied at leading order require improved theoretical precision through the inclusion of radiative corrections.
In this work, we present a fully differential Higgs plus one jet production at next-to-leading-order (NLO) accuracy in QCD for both the NC and CC processes. For the proposed Large Hadron electron Collider (LHeC) configuration, with a 60~GeV electron beam and a 7~TeV proton beam,
the total cross sections receive modest corrections with significantly reduced scale uncertainties. We find that in several kinematic distributions which are relevant to the analysis of $HVV$ couplings, the NLO K-factors are not flat.
Within the Standard Model, the polar angle of the electron (for NC) and the azimuthal angular correlation (for both NC and CC processes) receive maximum corrections in the range of 8-10\% in certain bins. We also compute NLO QCD corrections in the presence of non-standard $HVV$ interactions. The corrections in the azimuthal angular correlations are similar to the standard model predictions. For the polar angle of the electron, the corrections are sensitive to the nature of the $HVV$ coupling.

}

\begin{document}
\maketitle

\flushbottom

\newpage

%%%%%%%%%%%%%%%%%%%%%%%%%%%%%%%%%%%%%%%%%%%%%%%%%%%%%%%%%%%%%%
\section{Introduction}
\label{sec:intro}
Following the discovery of the Higgs boson ($H$) at the LHC~\cite{ATLAS:2012yve,CMS:2012qbp}, various Higgs production channels have been studied to determine its couplings to other particles in the Standard Model (SM). Precise measurements of the Higgs couplings to electroweak gauge bosons ($HVV$) are crucial in order to test the mechanism of electroweak symmetry breaking in the SM~\cite{Higgs:1964pj,Higgs:1964ia}. For this reason, vector boson fusion (VBF)—the second most prominent channel for single Higgs production—remains a key focus in LHC searches~\cite{Spira:2016ztx,LHCHiggsCrossSectionWorkingGroup:2016ypw}. The combined measurements by the CMS~\cite{CMS:2022dwd} and ATLAS~\cite{ATLAS:2022vkf} collaborations have achieved precision levels below 10\% for these couplings.  
Due to the cleaner environment with reduced QCD backgrounds and other advantages discussed in Refs.~\cite{LHeC:2020van,LHeCStudyGroup:2012zhm,Ahmadova:2025vzd}, the LHeC provides a complementary setting to the LHC. Thus, it can play a significant role in probing potential new physics directions in $HVV$ coupling.

The new physics can be parametrized by the most general Lorentz structure of two rank tensors made of momenta of vector bosons. The expression for the $HVV$ vertex is given by~\cite{Stong:1993bq,Dutta:2008bh,Biswal:2012mp,Alloul:2013naa,Kumar:2015kca,Hernandez-Juarez:2023dor}:
\begin{align}
	\Gamma_{HVV}^{\mu\nu}(p_1,p_2) = & g_V m_V \kappa_V g^{\mu\nu} - \frac{g}{m_W} [\lambda_{1V} (p_1^\nu p_2^\mu - g^{\mu \nu} p_1.p_2) \notag \\
& + \lambda_{2V} (p_1^{\mu}p_1^{\nu}+p_2^{\mu}p_2^{\nu}-g^{\mu \nu} p_1.p_1-g^{\mu \nu} p_2.p_2) \notag \\
& +  \widetilde{\lambda}_V ~ \epsilon^{\mu \nu \alpha \beta} p_{1 \alpha} p_{2 \beta} ].
	\label{Eq:BSMcoup}
\end{align}
Here, $p_1$ and $p_2$ are the momenta of the vector bosons. The BSM parameters $\kappa_V$ and $\lambda_{1V, 2V}$ are associated with CP-even couplings, while $\widetilde{\lambda}_V$ is associated with CP-odd coupling. $\epsilon^{\mu\rho\alpha\beta}$ is Levi-Civita tensor of rank 4 defined with $\epsilon^{0123}=1$ prescription. 

The non-standard $HVV$ couplings introduced in Eq.~\ref{Eq:BSMcoup} can be probed using observables sensitive to the Lorentz structure of the coupling. Several studies exploring these observables have been carried out at both hadron~\cite{Han:2009ra,Christensen:2010pf,Dwivedi:2016xwm,PhysRevD.83.113004,Plehn:2001nj,Hankele:2006ma} and lepton~\cite{Han:2000mi,Hagiwara:2000tk,Li:2019evl} colliders. This new coupling has also been studied in a few recent works ~\cite{Hernandez-Juarez:2024zpk,Hernandez-Juarez:2025nzh} in the context of the Higgs decays. In our previous studies~\cite{Sharma:2022epc,Sharma:2025ceq,Sharma:2024ujn}, we investigated non-standard couplings by analyzing the sensitivity of angular observables in single Higgs production via the VBF channel at the $ep$ collider. These studies estimated constraints on the BSM coupling parameters using angular observables computed at leading order (LO) in perturbation theory. However, higher-order corrections are essential to achieve a more precise understanding of these observables and of other kinematic distributions employed in selection cuts. Such corrections are expected to impact the bounds on BSM parameters derived from LO analyses.
Some results focusing on the impact of Next-to-leading order (NLO) QCD and electroweak corrections, within the SM framework, in NC and CC processes
 already exist in the literature~\cite{Blumlein:1992eh,Jager:2010zm,Wang:2022awk,Xiong:2023qnr}. 

The publicly available Monte Carlo packages~\cite{Sherpa:2024mfk,Alwall:2014hca} do not allow automated computation 
of QCD corrections in processes at the $ep$ collider in the presence of non-standard couplings.
In this study, we develop an in-house Monte Carlo code which incorporates $\mathcal{O}(\alpha_s)$ QCD corrections in the standard model as well as in the presence of non-standard $HVV$ couplings for the NC and CC processes at the $ep$ collider. Apart from revisiting the calculation for the inclusive cross sections,  we investigate the impact of corrections on angular observables sensitive to non-standard $HVV$ couplings. 

%%%%%%%%%%%%%%%%%%%%%%%%%%%%%%%%%%%%%%%%%%%%%%%%%%%%%%%%%%%%%% 

\section{The Process}
\label{sec:prcs}
\begin{figure}[!h]
  \begin{center}
\includegraphics [angle=0,width=0.5\linewidth]{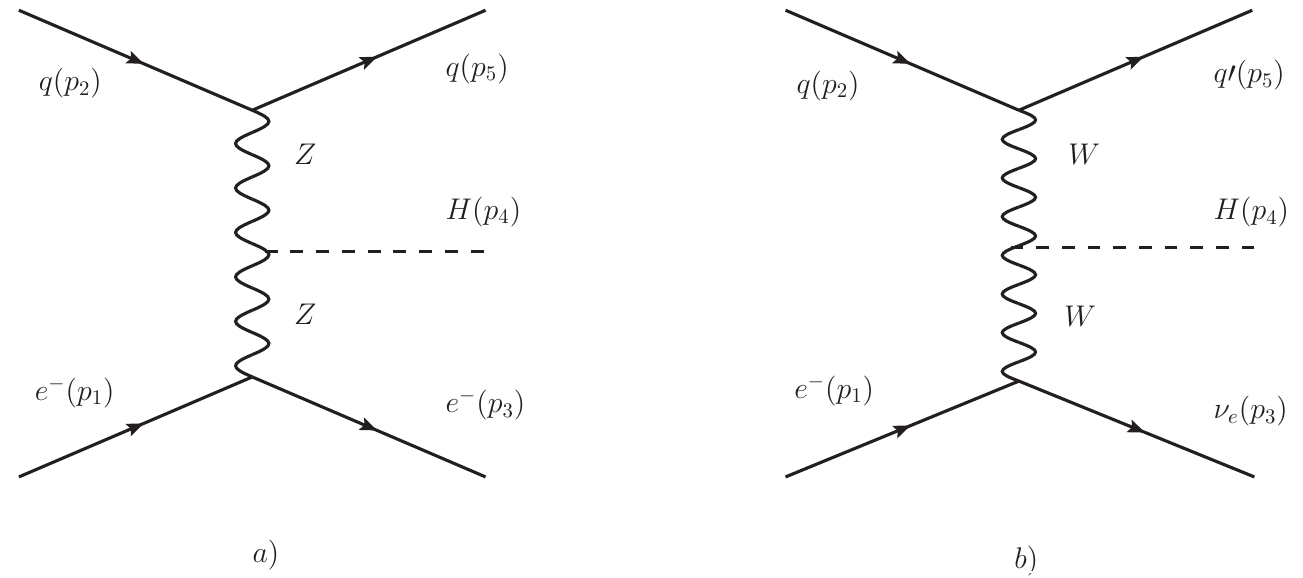}\\
	\caption{LO Feynman diagrams for NC and CC.  }
	\label{fig:tree_dia}
	\end{center}
\end{figure}

We are interested in computing NLO QCD corrections to $H+1$ jet production in $e^-p$ collision~\cite{Sharma:2022epc}.
%\citep{ATLAS:2022faz}
 At the leading order (LO), we are studying neutral current (NC) process $e^-p\rightarrow e^- H j$ and charged current (CC) process $e^-p\rightarrow \nu_e H j$. In the leading order (LO), there is only one tree-level Feynman diagram associated with each process.
As depicted in Fig.~\ref{fig:tree_dia}, diagram (a) is for the NC process and diagram (b) for the CC process. $HZZ$ vertex appears for NC current, and $HWW$ vertex appears for CC current process. 
  We are interested to see the new physics effects in these $HVV$ vertices.

\begin{figure}[!h]
  \begin{center}
\includegraphics [angle=0,width=0.5\linewidth]{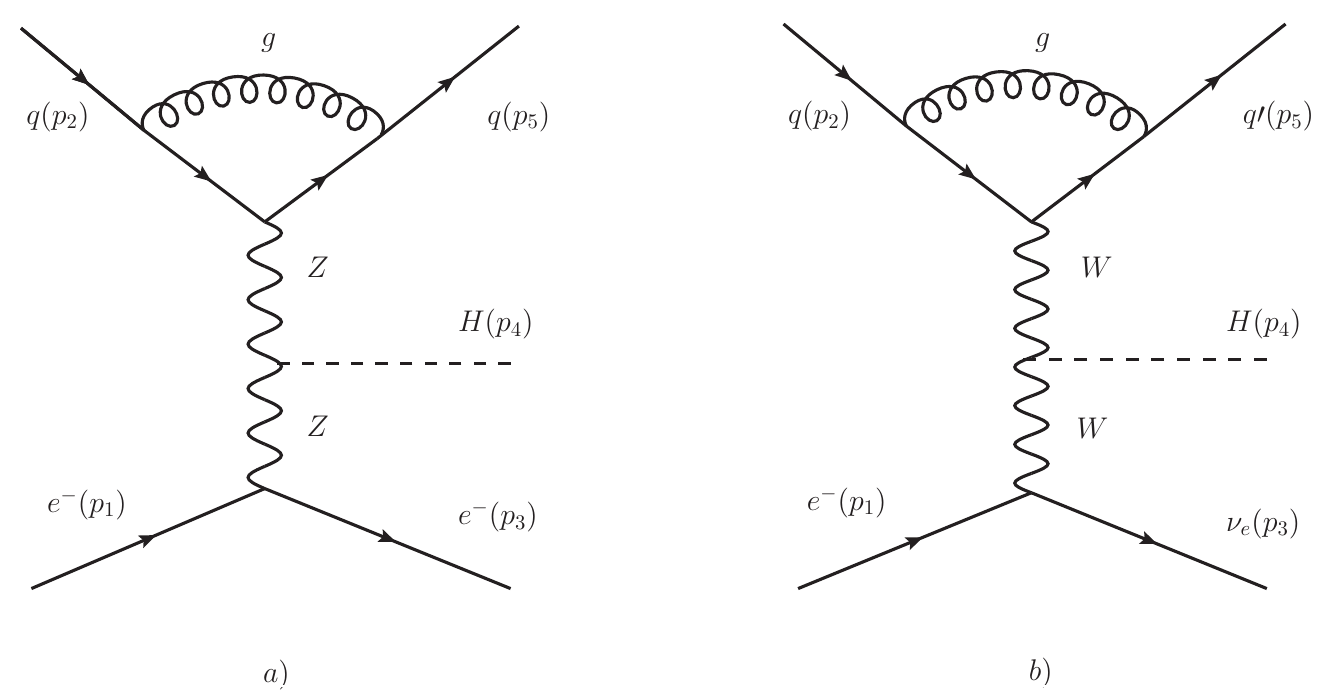}\\
	\caption{Virtual diagrams for NC and CC.  }
	\label{fig:virt_dia}
	\end{center}
\end{figure}

\begin{figure}[!h]
  \begin{center}
\includegraphics [angle=0,width=0.5\linewidth]{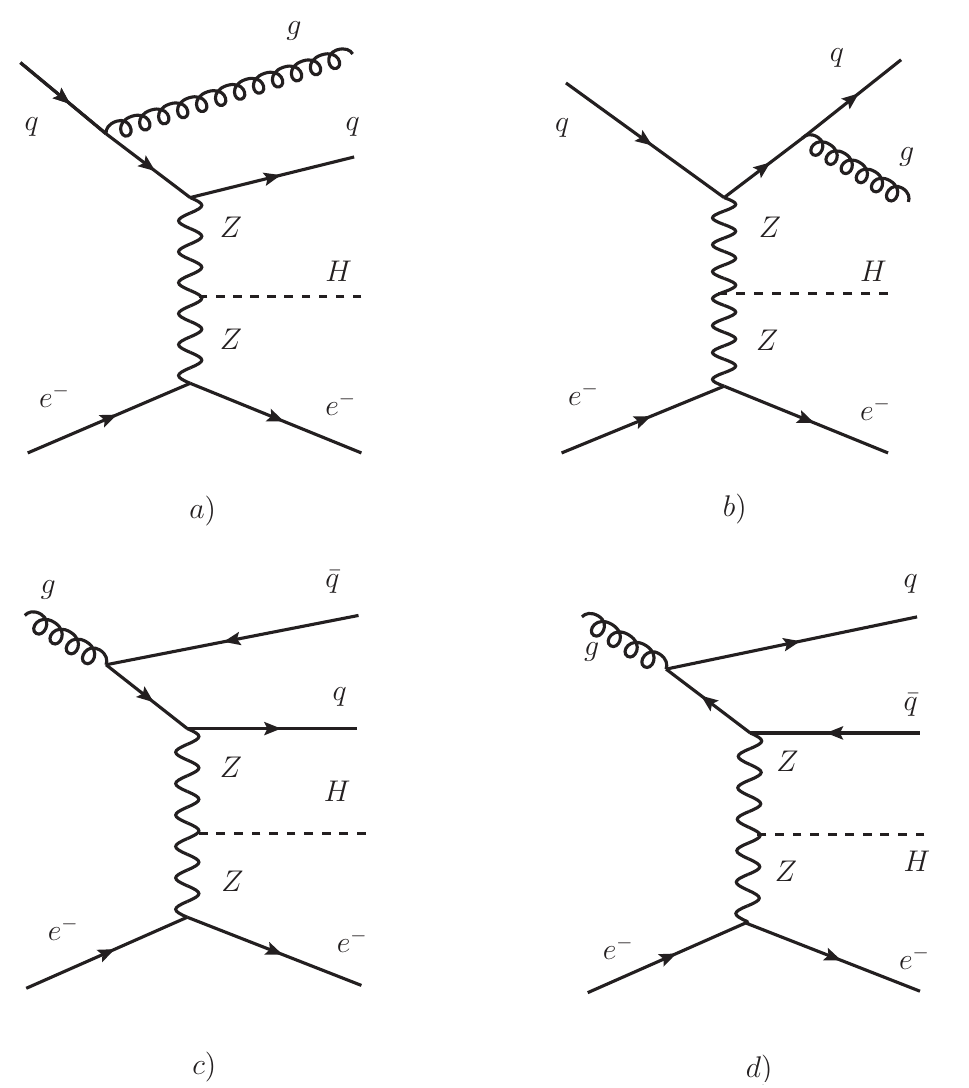}\\
	\caption{Real emission diagrams for NC.  }
	\label{fig:real_dia}
	\end{center}
\end{figure}

From Fig.~\ref{fig:tree_dia}, we see that there is one hadronic current and one leptonic current associated with tree-level diagrams. For this process, there is only one next-to-leading order diagram related to QCD vertex correction to the $qq\prime V$ vertex. The one-loop diagram is depicted in Fig.~\ref{fig:virt_dia}. For the inclusive observable, one also needs to consider the real emission diagrams. Gluons can be emitted from initial quarks as well as from final quarks.
There are also gluon-initiated processes, where a gluon can split into one quark pair and one of them will participate in hard scattering.
The inclusive processes are $e^-p\rightarrow e^- H jj$ for NC and $e^-p\rightarrow \nu_e H j j$ for the CC process. There are two sub-processes corresponding to each real emission process. The two sub-processes are quark-initiated and gluon-initiated processes. There are two real emission diagrams corresponding to each sub-process, as shown in Fig.~\ref{fig:real_dia}. We have shown real emission diagrams only for the NC process; similar types of diagrams exist for the CC process.

%%%%%%%%%%%%%%%%%%%%%%%%%%%%%%%%%%%%%%%%%%%%%%%%%%%%%%%%%%%%%%
\section{Calculations}
\label{sec:calc_check}
We calculate the helicity amplitudes for born, virtual, and real emission diagrams. As mentioned in the Sec.~\ref{sec:prcs}, there are one born, one one-loop, and four real emission (total) diagrams for each process (NC or CC).
 We take electron and light quarks as massless. The fermionic spinors have two helicity states, i.e, there can be $16$ helicity amplitudes for born and virtual diagrams. Because of the vector current structure, only $4$ of them will survive.
 In a similar fashion for real emission diagrams, out of $32$ amplitudes, only $8$ will survive. We use spinor helicity formalism to compute the helicity amplitudes~\cite{Peskin:2011in}. This formalism is being done for massless fermions. We calculate the helicity amplitudes in terms of scalar spinor products $\langle pq \rangle$ or $[pq]$~\cite{Peskin:2011in,KLEISS1986159,Agrawal:2021owa}.
   For real emission diagrams, there are gluons in the final states. The polarization of gluons has to be contracted with tensor currents at the vertex, such as with $\langle p\gamma^\mu q]$. We also compute the tensor currents from Dirac spinor solutions \cite{Das:2023gsh}. These currents follow all of the helicity identities in spinor helicity formalism.
  We compute helicity vectors' components numerically for massless gauge bosons.

  We calculate the virtual diagrams in the HV dimensional scheme~\cite{THOOFT1972189}. In this scheme, the loop part is being calculated in $d$-dimension and the rest is being calculated in $4$-dimension. We get the analytical expression for the loop diagram in the HV scheme. It can be written as
 \begin{equation}
 \mathcal{M}^V=-\frac{\alpha_s(\mu_R)}{2\pi}\frac{(4\pi)^\epsilon}{\Gamma(1-\epsilon)}C_F\Big(\frac{\mu^2}{2p_2.p_5}\Big)^\epsilon\Big\{-\frac{1}{\epsilon^2}-\frac{3}{2\epsilon}-4+\mathcal{O}(\epsilon)\Big\}\times \mathcal{M}^B.
 \label{equ:virt}
 \end{equation}

Here, $\mathcal{M}^V, \mathcal{M}^B$ stand for virtual- and born-level amplitudes. $\mu(\mu_R)$ is the scale (renormalization scale) and $\alpha_s(\mu_R)$ is the strong coupling at $\mu_R$. $C_F$ is the color factor which is $\frac{4}{3}$. $p_2$ and $p_5$ are the momenta of the initial and final quarks, respectively. The pole $\epsilon$ is defined as $(4-d)/2$. As we see in Eq.~\ref{equ:virt}, the virtual diagram has the highest pole $\frac{1}{\epsilon^2}$, which means the virtual diagram exhibits both soft and collinear divergences.
We know that the weak coupling $\alpha_{\text w}$ does not get any quantum correction at one-loop QCD correction, and wave function renormalization vanishes in on-shell renormalization for massless fermions. Here, we do not need any renormalization, and the $\frac{1}{\epsilon}$ pole in Eq.~\ref{equ:virt} is solely an IR pole.
The source of IR divergences and their cancellation with real emission diagrams have been discussed in Sec.~\ref{sec:ir_dp} in great detail.  

Once IR poles are removed from the renormalized amplitudes square, one has to perform the phase-space (PS) integrals for final state particles. Along with this, one has to perform an integral over the parton-distribution function (PDF) for each momentum fraction. As in $e^-p$ collision, one proton is taking part in the collision, there will be one PDF to integrate. 
 For the born and virtual diagrams, there are three body final state phase-space integrals, and for the real emission diagrams, there are four body final state phase-space integrals.
 We use {\tt RAMBOO}~\cite{Kleiss:1985gy} routine for phase-space points generators with given incoming momenta. The PDF and PS integral has been performed with the help of the Monte-Carlo integrator {\tt AMCI}~\cite{Veseli:1997hr}, which is based on {\tt VEGAS} algorithm~\cite{Lepage:1977sw}. {\tt AMCI} uses Parallel Virtual Machine {\tt PVM}~\cite{10.7551/mitpress/5712.001.0001} for parallel computation across the physical CPUs. We generate all of our necessary distributions with the help of the {\tt AMCI}.   The subtlety of getting distributions regarding dipole subtraction has been discussed in the following section.  

\section{IR singularities and Dipole subtraction}
\label{sec:ir_dp}

 The virtual diagrams in Fig.~\ref{fig:virt_dia} are IR divergent. As one can see from the figure, three massless particles are meeting at a vertex, and two on-shell particles are exchanging one massless particle, so the virtual diagrams are soft as well as collinear divergent \cite{Dittmaier:2003bc}.
 The IR divergent structure we can see from Eq.~\ref{equ:virt}. The poles in this equation are completely IR in nature. The real emission diagrams are also IR singular in soft and collinear regions due to the extra fermionic propagator, which splits into gluon and fermion.
One can also show that real emission diagrams can exhibit the same pole structure as Eq.~\ref{equ:virt} if the phase-space for extra emission is integrated over in $d=4-2\epsilon$ dimensions.
  Due to the collider search in fiducial regions and the complicated structure of the matrix element, we integrate the phase space over the emitted parton numerically. Also, we are interested in several kinematical distributions, which motivate us to develop a Monte-Carlo event generator for LO and NLO contributions.

 We adopt Catany-Seymour dipole subtraction~\cite{Catani:1996vz} to remove IR singularities from the inclusive partonic cross-section and to study inclusive differential distributions. The idea is to introduce a local counter-term which will absorb singularity from virtual amplitudes and at the same time exhibit point-wise singularities as the real emission amplitudes. The total inclusive cross-section can be written as 
 
\begin{equation}
\sigma^{\text{NLO}}=\int_{m+1} d\sigma^{\text R} + \int_{m} d\sigma^{\text V}
\label{equ:nlo_v_r}
\end{equation}  
 
Following the Ref.~\cite{Catani:1996vz}, we rewrite this Eq. as
\begin{equation}
\sigma^{\text{NLO}}=\int_{m+1} d\sigma^{\text R}-\int_{m+1} d\sigma^{\text A} + \int_{m} d\sigma^{\text V}+\int_{m+1} d\sigma^{\text A}.
\label{equ:nlo_v_r_a}
\end{equation}
The local counterterm $\delta \sigma^A$ has the same pointwise singular behavior as real emission diagrams in soft and collinear regions. Integrating over extra phase-space in $\delta \sigma^A$ will remove all of the IR singularities from virtual diagrams. 
 We can write 
 \begin{equation}
 \sigma^{\text{NLO}}=\int_{m+1} [d\sigma^{\text R}- d\sigma^{\text A}] + \int_{m} \Big[d\sigma^{\text V}+\int_{1} d\sigma^{\text A}\Big].
 \end{equation}
 The last term in the second integral in the above equation will lead to the insertion operator {\textit {\textbf I}} which has to be convoluted with born level cross section. In this process, there is one identified parton in the initial state, and there is a finite remainder which is left after the factorization of initial state collinear singularities into the non-perturbative parton distribution functions (PDF). These terms are called {\textit {\textbf P}}, {\textit {\textbf K}} terms. Following the Ref.~\cite{Catani:1996vz}, we write the final dipole-subtracted NLO cross section as
 \begin{eqnarray}
 \sigma^{\rm NLO}(p)=&&\int_{m+1} \Big[\big(d\sigma^R(p)\big)_{\epsilon=0}-\Big(\sum_{\text{dipoles}}d\sigma^{\text B}(p)\otimes (dV_{\text{dipoles}}+dV^\prime_{\text{dipoles}})\Big)_{\epsilon=0}\Big]\\ \nonumber
&&+\int_m \big[d\sigma^{\text V}(p)+d\sigma^{\text B}(p) \otimes {\textit {\textbf I}} \big ]_{\epsilon=0} + \int_0^1 dx \int_m \big[d\sigma^{\text B}(xp)\otimes ({\textit {\textbf P}} + {\textit {\textbf K}})\big]_{\epsilon=0}.
 \end{eqnarray}
Where $p$ is the momentum of the identified parton and $x$ is the longitudinal fraction of $p$.
 
The initial and final state partons are quarks for these processes. The insertion operator {\textit {\textbf I}} which has been computed in $d$-dimensions, can be written as 

\begin{equation}
{\textit {\textbf I}}(\epsilon)=\frac{\alpha(\mu_R)}{2\pi}\frac{(4\pi)^\epsilon}{\Gamma(1-\epsilon)}2C_F\Big(\frac{\mu}{2p_2.p_5}\Big)^\epsilon \Big\{\frac{1}{\epsilon^2}+\frac{3}{2\epsilon}-\frac{\pi^2}{2}+5+\mathcal{O}(\epsilon)\Big\}
\end{equation}
 
  For the processes under consideration, there is one initial state parton and one final state parton. So, only two dipole terms are needed. The dipole $\mathcal{D}_{ij}^a$ is for when the final state parton is the emitter and the initial state is the spectator, and the dipole  $\mathcal{D}_{k}^{ai}$ is for when the initial state is the emitter and the final state is the spectator. These two dipoles, along with the real emission amplitude square, will project all singular numbers to regular numbers in singular regions. 
  
  The finite collinear remainder is
  
  \begin{eqnarray}
  \sigma^{\rm NLO,c}(\mu_R,\mu_F)=\sum_a \int_0^1 dx_a f_q^c(x_a, \mu_R^2,\mu_F^2)\hat{\sigma}^{\text{LO}}_q(x_aP,p_e;\{p_f\})
  \end{eqnarray}
 where we define $f^c_q(x,\mu_F,\mu_R)$ as

\begin{eqnarray}
 f^c_q(x,\mu_F,\mu_R)=\frac{\alpha_s(\mu_R)}{2\pi}\Big[\int_x^1 \frac{dz}{z}\Big\{f_g(x/z,\mu_F^2)A(z)+f_q(x/z,\mu_F^2)B(z)\\ \nonumber
 +zf_q(x)C(z)\Big\}+f_q(x)D(x)\Big].
 \end{eqnarray}

Following Ref.~\cite{Catani:1996vz}, we compute the integral kernels A, B, C, and D. They can be written as

\begin{eqnarray}
A(z)&&=T_R\Big\{\Big(z^2+(1-z)^2\Big)\Big(\text{ln}\frac{1-z}{z}-\text{ln}\frac{\mu_F^2}{2p_2.p_5}\Big)+2z(1-z)\Big\}\\
B(z)&&=C_F\Big\{\frac{2\text{ln}(1-z)}{1-z}-2\frac{\text{ln}z}{1-z}-(1-z)\text{ln}\frac{1-z}{z}+(1-z)\\ \nonumber
&&\quad\quad\quad\quad\quad\quad\quad\quad\quad\quad-\frac{3}{2}\frac{1}{1-z}-\frac{1+z^2}{1-z}\text{ln}\frac{\mu_F^2}{2p_2.p_5}\Big\}\\ 
C(z)&&=C_F\Big\{-\frac{2\text{ln}(1-z)}{1-z}+\frac{3}{2}\frac{1}{1-z}+\frac{2}{1-z}\text{ln}\frac{\mu_F^2}{2p_2.p_5}\Big\}\\
D(x)&&=C_F\Big\{\text{ln}^2(1-x)-\frac{\pi^2}{3}-5+\pi^2-\frac{3}{2}\text{ln}(1-x)-\frac{3}{2}\\ \nonumber
&&\quad\quad\quad\quad\quad\quad\quad\quad\quad-\big(2\text {ln}(1-x)+\frac{3}{2}\big)\text{ln}\frac{\mu_F^2}{2p_2.p_5}\Big\}
\end{eqnarray}

    where $T_R=\dfrac{1}{2}$ . These kernels have been consistent with the Ref.~\cite{Jager:2010zm}

\section{Checks}
\label{sec:num_res}

We check our calculations and implementation in several ways to avoid any subtle mistakes and ensure the reliability of our computations. 
Born level cross-sections for both NC and CC processes are in agreement with the {\tt MadGraph}~\cite{Alwall:2014hca} results with the same set of SM parameters, PDF, and cuts.
We have made the Ward identity check on the real amplitudes by replacing the polarization of the external gluon with its momentum.
The real contribution is also in good agreement with the {\tt MadGraph} with similar cuts on radiated partons.
Virtual diagram calculation--pole structures, finite pieces match with the reference Ref.~\cite{Jager:2010zm}.
IR poles cancel with the insertion operator {\textbf {\textit I}}.
The dipole contribution approaches the real contribution in soft and collinear regions, making the net contribution (real-dipole) finite. This validates the correct implementation of the CS dipole subtraction scheme.

\section{ Numerical Results}
\label{sec:num_res}
 We construct jets out of the final state partons to probe in fiducial regions and kinematical distributions. The tree level and virtual diagrams have only one parton in the final state, so there can be at most one jet production associated with the Higgs. This is the case with collinear remainder and dipole terms, as these contributions are convoluted with born contribution. The real emission diagrams have two partons in the final state, so there can be a maximum of two jets produced associated with the Higgs.
 The jets are constructed from small-angle partons within the distance $R=0.4$ in $(\eta-\phi)$ plane. We use {\tt FastJet}~\cite{Cacciari:2011ma,Cacciari:2005hq} package for jet clustering. We impose a minimum $20$ GeV and $10$ GeV cut on transverse momenta ($p_T$) of jets and electron (missing $E_T$ for neutrinos), respectively. We also separate jets and the electron, but imposing a distance cut $\Delta R=0.4$ in $(\eta-\phi)$ plane.
 The born, virtual, collinear kernels and dipoles are evaluated with three-body final state Monte-Carlo phase-space integration, whereas the real emission contribution has been evaluated with four-body final state Monte-Carlo phase-space integration. 
 The kinematical cuts have to be implemented separately on real emissions and dipole contributions. The CS dipole subtraction ensures the cancellation of these two contributions in singular regions. 
  
We use the SM parameters set from PDG2024~\cite{ParticleDataGroup:2024cfk}. The masses and widths of vector bosons are taken as $M_W=80.3692$ GeV, $M_Z=91.188$ GeV, $\Gamma_W=2.085$ GeV, and $\Gamma_Z=2.4955$ GeV. The Higgs mass is taken as $M_H=125$ GeV. We use $\alpha_{W}(M_Z)=1/127.93$ in $\overline{MS}$ scheme. We choose a dynamical scale choice for renormalization and factorization scale as

\begin{equation}
\mu_0=\mu_R=\mu_F=\frac{1}{3}\Big(p_{T;j}+\sqrt{M_H^2+p_{T;H}^2}+p_{T;e^-/\nu_e}\Big),
\end{equation}  
where $p_T$ is the transverse momentum of the particle. We vary this scale in the range $0.5\mu_0\geq \mu \geq 2\mu_0$ to measure the scale uncertainties. We use the {\tt LHAPDF}~\cite{Buckley:2014ana} interface routine to calculate the strong coupling $\alpha_s(\mu_R)$ for each phase-space point. In the following, we define the \textit{relative increment} at the level of total cross section due to NLO QCD correction as
\begin{equation*}
\text{RI} = \frac{\sigma^{\text{NLO}} - \sigma^{\text{LO}}}{\sigma^{\text{LO}}} \times 100\% = (K-1) \times 100\%.
\end{equation*}
In the above, $K=\sigma_{\rm NLO}/\sigma_{\rm LO}$ is defined as the NLO K-factor. We will use this quantity to discuss the QCD corrections in different bins of various kinematic and angular distributions. 

\subsection{SM prediction}
\label{subsec:sm_pdct}
First, we report NLO corrections to the total cross sections for the neutral current NC and CC processes, which have been computed using the kinematic cuts outlined at the beginning of this section. In Tab.~\ref{table:sm_results}, we 
present LO and NLO cross sections using two PDF sets \texttt{CT18}~\cite{Guzzi:2022rca,Hou:2019efy} and \texttt{NNPDF40}~\cite{NNPDF:2021njg}. With the \texttt{CT18} PDF, the SM cross-sections at LO and NLO for the NC process are $17.44~\text{pb}$ and $18.14~\text{pb}$, respectively, while for the CC process they are $91.86~\text{pb}$ and $96.12~\text{pb}$. 
 We find positive NLO corrections, with the RI amounting to $4.01\%$ for the NC process and $4.64\%$ for the CC process.
The CC process exhibits a slightly larger correction compared to the NC process. In the table, the superscript and subscript to total cross-sections are scale uncertainties corresponding to $\mu = 0.5\mu_0$ and  $ \mu =
 2\mu_0$, respectively. At LO, only the factorization scale $\mu_F$ enters through the PDF, while the renormalization scale $\mu_R$ appears at NLO through the strong coupling constant $\alpha_s(\mu_R)$. Therefore, the scale uncertainty at LO arises solely from variations in $\mu_F$, and this is around $\pm5$\% for the NC and CC processes. At NLO, a partial cancellation of the $\mu_F$ dependence occurs between the PDF and the partonic cross section. Additionally, a relatively small size of the loop corrections adds a mild dependence on $\mu_R$. Consequently, the overall scale uncertainty is reduced below 1\% when going from LO to NLO. We also observe that the scale uncertainties are identical for both the NC and CC processes at each perturbative order. This is expected due to diagrammatic similarity between the two processes.

\newcolumntype{M}[1]{>{\centering\arraybackslash}m{#1}}
\renewcommand{\arraystretch}{1.9}
\begin{table}[H]
\begin{center}
\begin{tabular}{c|c|c|c|c}
\hline
Process& PDF set &$\sigma^{\rm LO}$(pb)&$\sigma^{\rm NLO}$(pb)&RI($\%$)\\
\hline
 \multirow{2}{*}{NC} 
			& CT18 &  $17.44^{+5.27\%}_{-4.61\%}$ &$18.14^{-0.08\%}_{-0.52\%}$&4.01\\
         &  NNPDF40 & $19.57^{+4.82\%}_{-4.25\%}$ &$18.44^{-0.02\%}_{-0.48\%}$&-5.77\\
\hline
\multirow{2}{*}{CC}& CT18 &$91.86^{+5.29\%}_{-4.59\%}$&$96.12^{+0.06\%}_{-0.48\%}$& 4.64\\
& NNPDF40 & $100.89^{+4.74\%}_{-4.18\%}$ & $96.78^{+0.06\%}_{-0.47\%}$ & -4.07\\
\hline
\end{tabular}
\caption{SM cross-sections and RIs for NLO QCD corrections in the NC and CC processes for \texttt{CT18} and \texttt{NNPDF40} PDF sets. The numbers in the subscript and superscript denote the percentage scale uncertainties.}
\label{table:sm_results}
\end{center}
\end{table}

For \texttt{NNPDF40} PDF set, the LO and NLO cross-sections are 19.57 (100.89) pb and 18.44 (96.78) pb for the NC (CC) process.
Contrary to the \texttt{CT18} PDF choice, for the \texttt{NNPDF40} PDF set, we get negative corrections with RI -5.77\% and -4.07\% in the NC and CC processes, respectively.  
Scale uncertainties for both PDF sets are similar for LO and NLO cross-sections. For NC (CC), although the LO cross-section varies by $\pm$11.51\% ($\pm$9.37\%), the NLO cross-section varies by $\pm$1.64\% ($\pm$0.68\%) for these two PDF sets. The variation (band) of the PDF set uncertainty can be estimated by calculating the cross-sections with several PDF sets.
 In this work, we present the rest of the results using \texttt{CT18} PDF set only.

\begin{figure}[htp]
\begin{center}
\includegraphics [angle=0,width=0.37\linewidth]{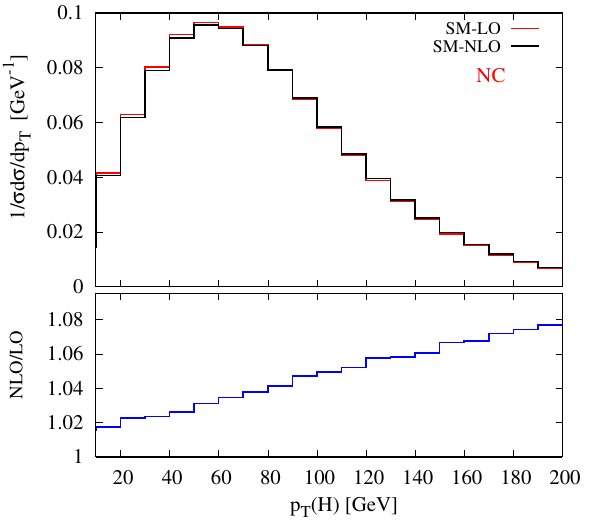}
\includegraphics [angle=0,width=0.37\linewidth]{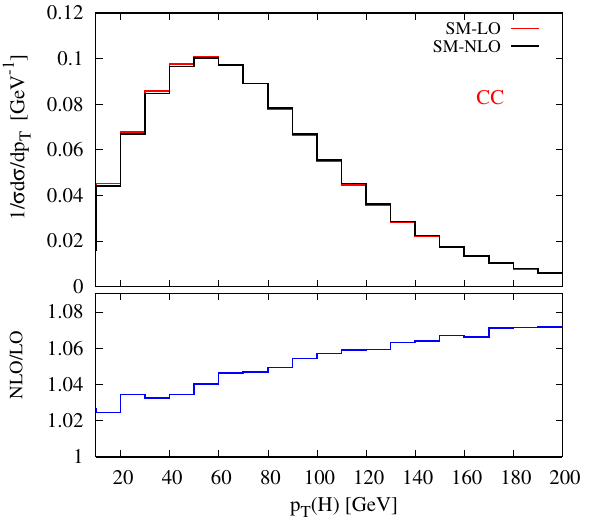}
\includegraphics [angle=0,width=0.37\linewidth]{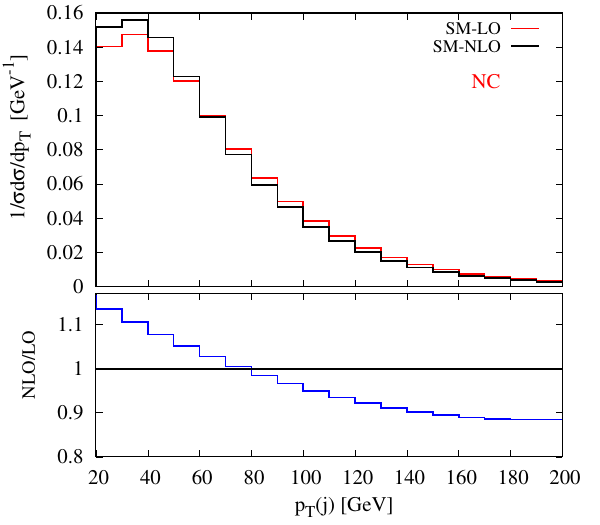}
\includegraphics [angle=0,width=0.37\linewidth]{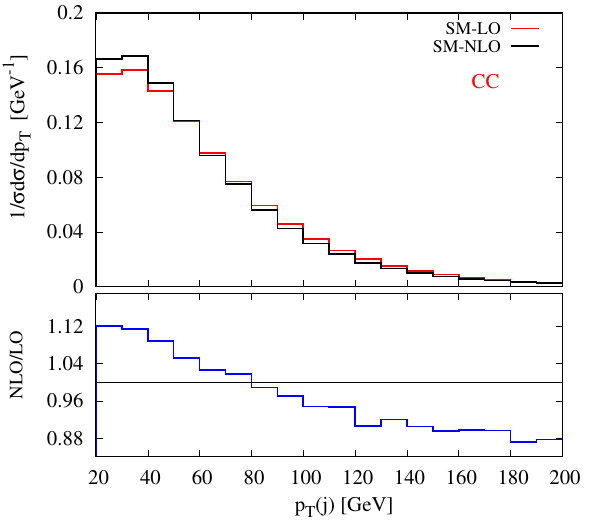}
\includegraphics [angle=0,width=0.37\linewidth]{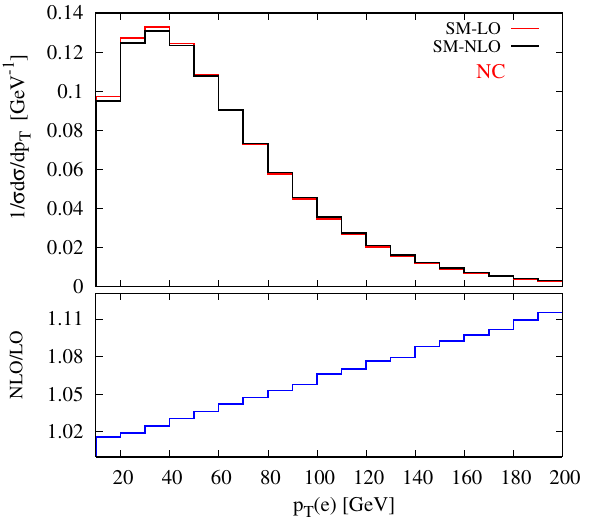}
\includegraphics [angle=0,width=0.37\linewidth]{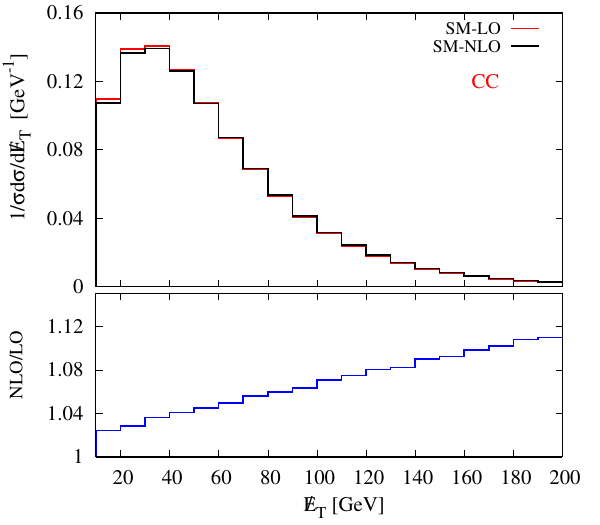}
\caption{ Upper panel: LO (red) and NLO QCD (black) predictions for $p_T$ distribution of final state particles in the NC (left) and CC (right) processes with SM couplings. Lower panel: NLO K-factor (blue) 
in different bins of the $p_T$ distributions. %and invariant masses ($M_{ij/ijk}$) for NC.
}
\label{fig:nc_sm_pt}
\end{center}
\end{figure}

Now we discuss the effect of NLO corrections on the differential distributions essential for selection cuts to optimize the signal over background. The normalized differential distributions are presented along with the corresponding K-factors. Since the corrections from dipole-subtracted real amplitudes contribute significantly to the total correction, the K-factors for certain distributions exhibit fluctuations. In such cases, we fit the K-factors using polynomial functions that respect the overall behavior of the distributions.
In Fig.~\ref{fig:nc_sm_pt}, we show the normalized transverse momentum ($p_T$) distributions, i.e., $\frac{1}{\sigma}\frac{d\sigma}{d p_T}$, for key final-state objects. These include the tagged jet and the Higgs boson in both the NC and CC processes. Additionally, we present the $p_T$ distribution of the outgoing $e^-$ in the NC channel and the missing transverse energy ($\slashed{E}_T$), arising from the neutrino, in the CC channel. We observe that the $p_T$ distributions for each final state in both the NC and CC processes exhibit a similar pattern of NLO corrections. 
For the Higgs boson, the correction is 2\% at low $p_T$ and gradually rises with increasing $p_T$, reaching up to 8\%. In case of the tagged jet, it receives corrections of up to 12\% at low $p_T$, which become negative at higher $p_T$, reaching up to -12\% in the tail of the distribution. In particular, the behavior of the NLO corrections in the $p_T$ distribution for the jet closely follows the behavior reported in Ref.~\cite{Jager:2010zm}. The NLO corrections to the $p_T$ distribution of the $e^-$ in the NC process and the missing transverse energy ($\slashed{E}_T$) in the CC process increase progressively toward the high-$p_T$ region. For the electron and $\slashed{E}_T$, the corrections reach up to approximately 11\%.

\begin{figure}[htp]
\begin{center} 
\includegraphics [angle=0,width=0.37\linewidth]{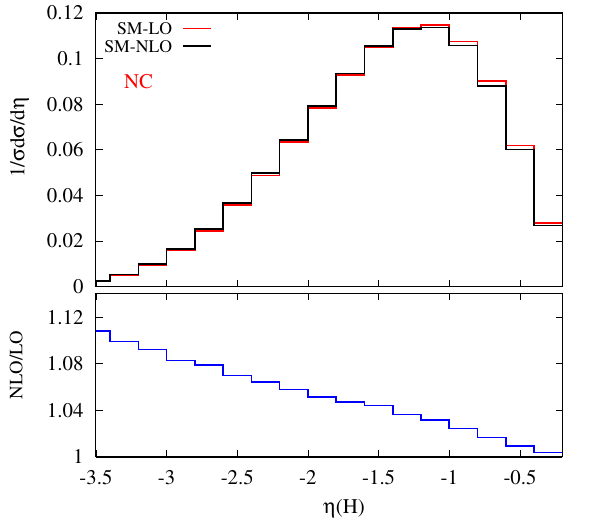}
\includegraphics [angle=0,width=0.37\linewidth]{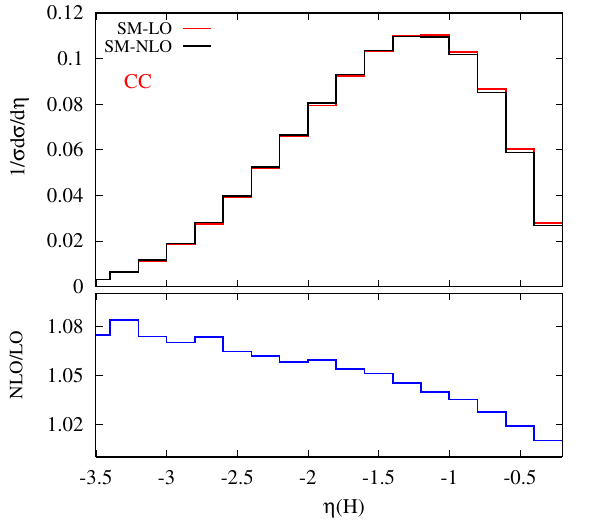}
\includegraphics [angle=0,width=0.37\linewidth]{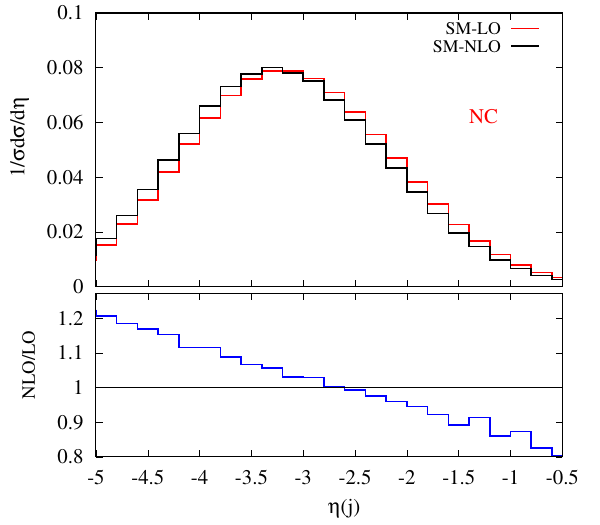}
\includegraphics [angle=0,width=0.37\linewidth]{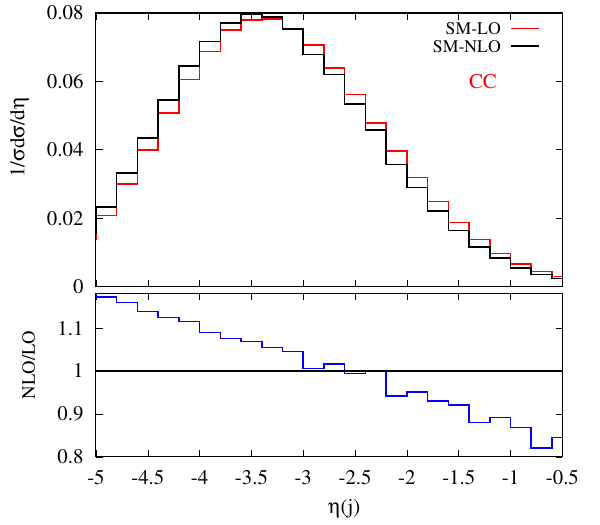}
\includegraphics [angle=0,width=0.37\linewidth]{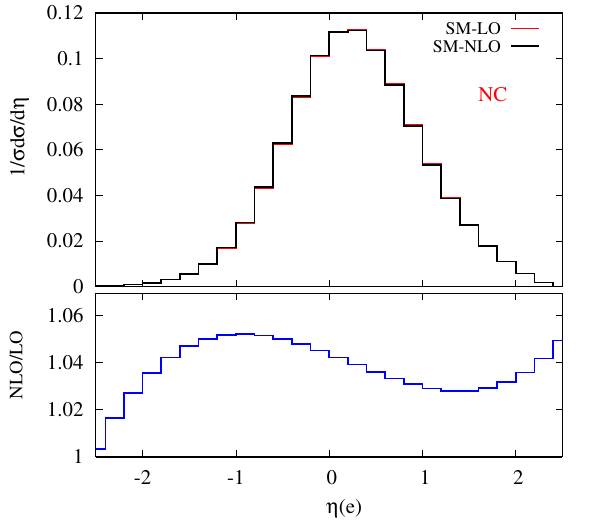}
\caption{ Upper panel: LO (red) and NLO QCD (black) predictions for pseudo-rapidities of observed final state particles in the NC (left) and CC (right) processes with SM couplings. Lower panel: NLO K-factor (blue) 
in different bins of the $\eta$ distributions. 
}
\label{fig:nc_sm_eta}
\end{center}
\end{figure}

The normalized pseudo-rapidity distributions ($\frac{1}{\sigma}\frac{d\sigma}{d\eta}$), along with the corresponding $K$-factors, for the final-state particles in the NC and CC processes are shown in Fig.~\ref{fig:nc_sm_eta}. In both processes, the NLO corrections lead to a noticeable shift in the $\eta$ distributions of the Higgs boson and the tagged jet toward higher rapidity. In the case of the tagged jet, this behavior can be attributed to the emission of an additional parton at NLO.
 The NLO corrections to the pseudo-rapidity distribution of the Higgs boson reach up to approximately 11\% in the NC process and 8\% in the CC process. However, the corrections for the tagged jet are larger, varying between -20\% to 20\% in the NC channel, and between -20\% to 18\% in the CC channel. The $\eta$ distribution for the electron in the NC process has a correction maximum up to 5\% in the region around $\eta=-1$.
The behavior of the NLO correction in the $p_T$ distribution for the scattered $e^-$ (and jet) and the $\eta$ distribution for the jet is consistent with the findings reported in Ref.~\cite{Jager:2010zm}.

\begin{figure}[htp]
\begin{center} 
\includegraphics [angle=0,width=0.37\linewidth]{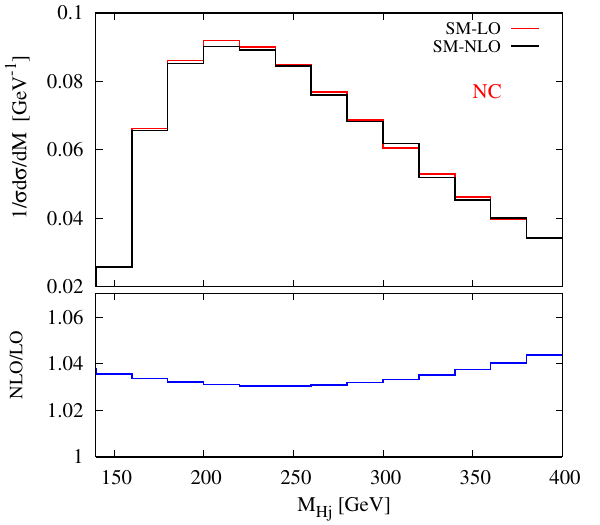}
\includegraphics [angle=0,width=0.37\linewidth]{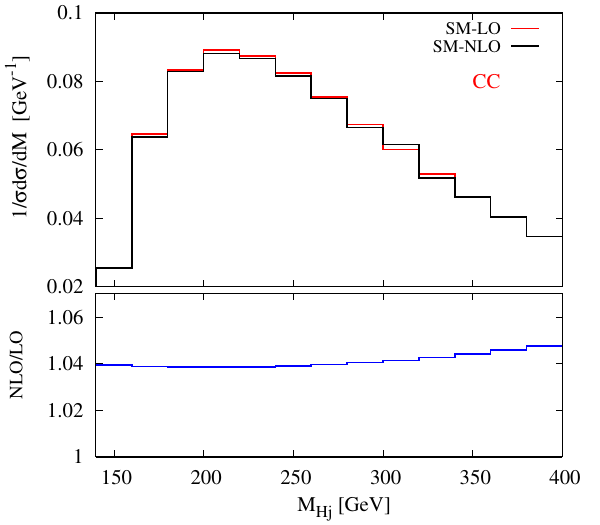}
\caption{ Upper panel: LO (red) and NLO QCD (black) predictions for the Higgs-jet invariant mass ($M_{Hj}$) distribution in the NC (left) and CC (right) processes with SM couplings. Lower panel: NLO K-factor (blue) 
in different bins of the $M_{Hj}$ distributions.   
}
\label{fig:sm_InMHj}
\end{center}
\end{figure}

The invariant mass distribution of the Higgs boson and the leading jet ($M_{Hj}$) plays a critical role in optimizing the signal-to-background ratio. Consequently, understanding the impact of QCD corrections on this distribution is essential. In Fig.~\ref{fig:sm_InMHj}, we present the impact of QCD corrections on the $M_{Hj}$ distribution for both the NC and CC processes. The corrections exhibit a similar trend in both cases, remaining approximately flat at around 4\% across all bins.

\begin{figure}[htp]
\begin{center}
\includegraphics [angle=0,width=0.37\linewidth]{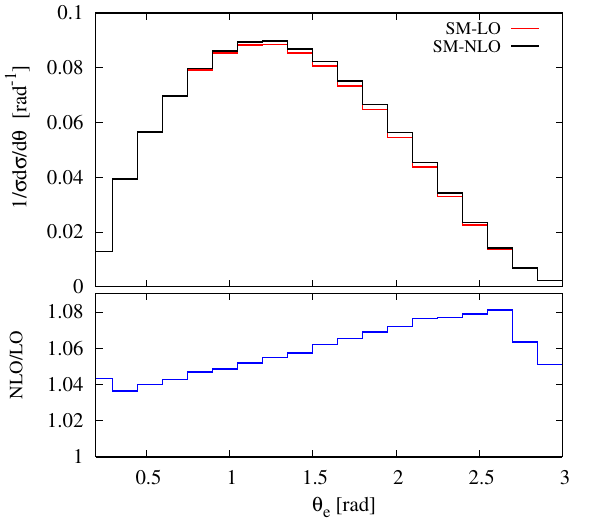}
\includegraphics [angle=0,width=0.37\linewidth]{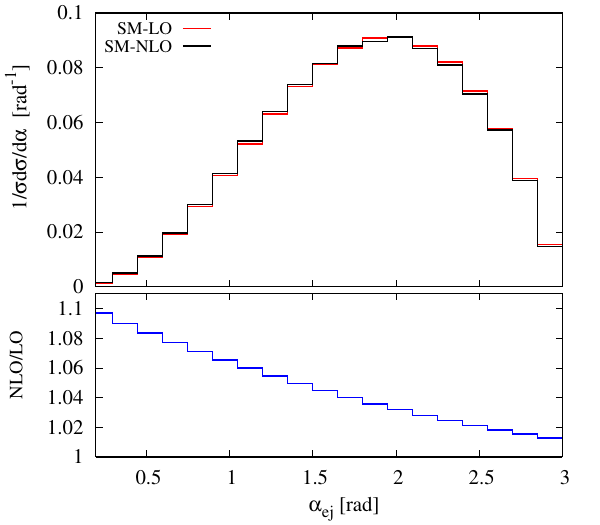}
\includegraphics [angle=0,width=0.37\linewidth]{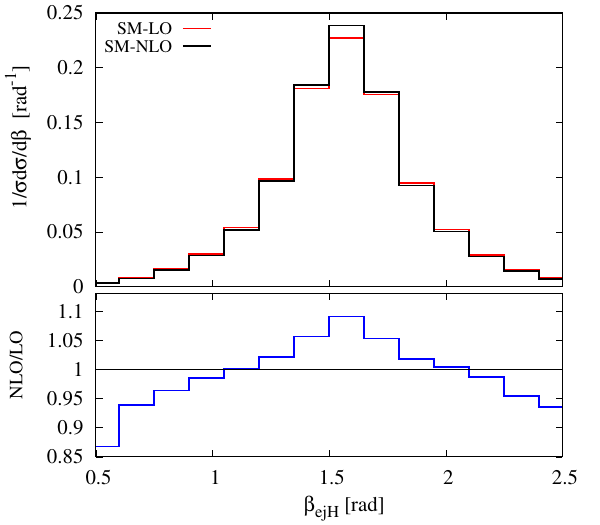}
\caption{ The LO (red) and NLO (black) predictions for angular variables $\theta_e$, $\alpha_{ej}$, and $\beta_{ejH}$ distributions for NC in SM. The bin-wise NLO K-factors are shown in the lower panels.}
\label{fig:nc_sm}
\end{center}
\end{figure}

The angular observables that provide the most stringent constraints on the $HVV$ coupling are discussed in Ref.~\cite{Sharma:2022epc,Sharma:2025ceq,Sharma:2024ujn} in detail. In the NC process, these observables include $\theta_e$, the polar angle of the scattered electron; $\alpha_{ej}$, the angle between scattered $e^-$ and the jet; $\beta_{ejH}$, the angle between Higgs and normal to the plane of $e^-$ and jet; and  $\Delta \phi_{ej}$, the azimuthal correlation of the electron and the jet, which ranges from $-\pi$ to $\pi$.  In case of the CC process, $\Delta \phi_{\slashed{E}j}$ distributions are relevant for the discussion because only information in the transverse plane is accessible due to missing energy in this process.  These observables have been studied at LO in perturbation theory and require higher-order corrections to capture their behavior more precisely, thereby potentially affecting the constraints on the $HVV$ coupling parameters.

\begin{figure}[htp]
\begin{center}
\includegraphics [angle=0,width=0.37\linewidth]{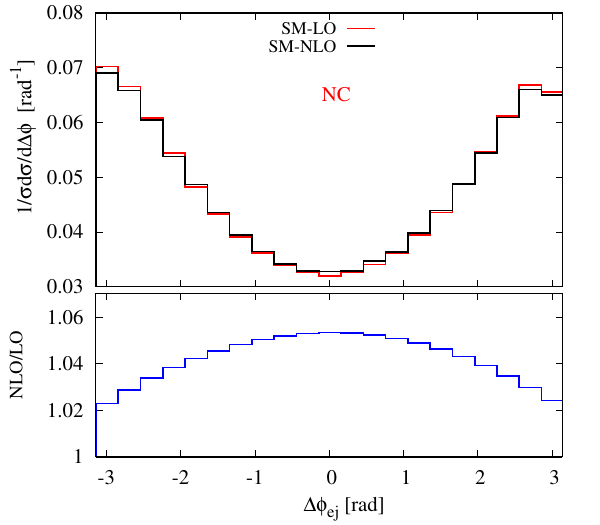}
\includegraphics [angle=0,width=0.37\linewidth]{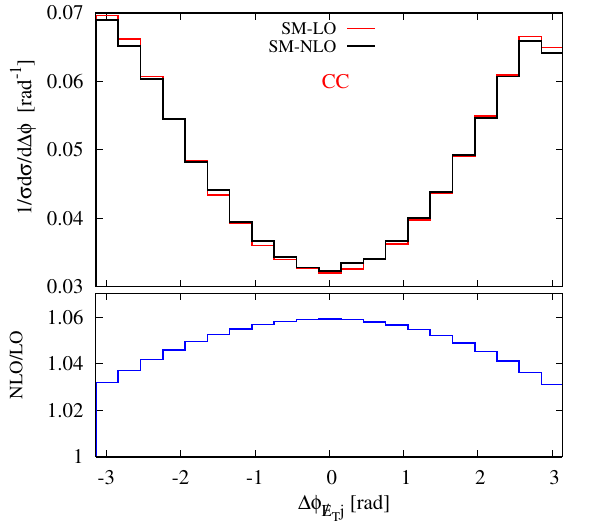}
\caption{ The LO (red) and NLO QCD (black) predictions for the angular variables $\Delta \phi $ distribution,  for NC (left) and CC (right) in SM. The bin-wise NLO K-factors are shown in the lower panels.}
\label{fig:norm_dphi}
\end{center}
\end{figure}

The $\theta_e$, $\alpha_{ej}$, and $\beta_{ejH}$ distributions associated with the NC process are presented  in Fig.~\ref{fig:nc_sm}. The $\theta_e$ and $\alpha_{ej}$ distributions are correlated, and consequently, their NLO corrections exhibit a correlated pattern as well. The $\theta_e$ distribution receives corrections in the range of 4–8\%, while those for $\alpha_{ej}$ reach up to 10\%. The $\beta_{ejH}$ is a CP-sensitive observable which is symmetric about $\pi/2$~\cite{Sharma:2025ceq}. Due to the CP-even nature of the QCD interactions, the NLO corrections are also symmetric, with positive shifts reaching up to 9\% in the central region and -13\% in the outer bins. In Fig.~\ref{fig:norm_dphi}, we present the $\Delta\phi_{ej}$ distribution for the NC process and the $\Delta\phi_{\slashed{E}j}$ distribution for the CC process. Similar to $\beta_{ejH}$, these observables display a symmetric correction pattern around the central bin. This is expected as $\beta_{ejH}$ and $\Delta \phi_{ej}$ are correlated and differ mainly in normalization~\cite{Sharma:2025ceq}. 
Notably, the NLO corrections to $\Delta\phi_{ej}$ and $\Delta\phi_{\slashed{E}j}$ are comparable in both magnitude and behavior. The corrections are around 5-6\% in the mid region of both distributions.

\subsection{New physics results}

The NC and CC processes involving non-standard $HVV$ couplings also receive NLO QCD corrections on the $q q^\prime V$ vertex, similar to the purely SM scenario. We keep one BSM parameter associated with each component of $HVV$ coupling non-zero at a time. We choose benchmark values for the $HZZ$ parameters as $\lambda_{1Z} = 0.4$, $\lambda_{2Z} = 0.3$, and $\widetilde{\lambda}_Z = 0.5$.  For the $HWW$ coupling, the benchmark values are $\lambda_{1W} = 0.07$, $\lambda_{2W} = 0.02$, and $\widetilde{\lambda}_W = 0.2$. These values are motivated by constraints obtained from a cross-section level $\chi^2$ analysis described in the Ref.~\cite{Sharma:2022epc}.

\newcolumntype{M}[1]{>{\centering\arraybackslash}m{#1}}
\renewcommand{\arraystretch}{1.7}
\begin{table}[htp]
\begin{center}

\begin{tabular}{c|c|c|c|c}
			\hline
			\multicolumn{2}{c|}{Process}  &$\sigma^{\rm LO}$(pb)&$\sigma^{\rm NLO}$(pb)&RI($\%$)\\
			\hline	
            \multirow{3}{*}{NC} 
			& $\lambda_{1Z}$ & $29.22^{+5.27\%}_{-4.61\%}$ & $30.08_{-0.38\%}^{-0.22\%}$ & 2.94\\
         &  $\lambda_{2Z}$ & $13.70^{+6.71\%}_{-5.75\%} $ &$14.26^{-0.03\%}_{-0.69\%}$&4.09\\
         & $\widetilde{\lambda}_Z$ & $25.25^{+5.46\%}_{-4.77\%}$ &$26.14^{-0.20\%}_{-0.53\%}$&3.52\\
         \hline
            \multirow{3}{*}{CC} 
			& $\lambda_{1W}$ & $99.49^{+5.23\%}_{-4.55\%}$ &$103.89^{-0.07\%}_{-0.48\%}$&4.42\\
			& $\lambda_{2W}$ & $82.20^{+5.25\%}_{-4.54\%}$ &$86.05^{+0.03\%}_{-0.50\%}$ &4.68\\
            &  $\widetilde{\lambda}_W$ & $102.39^{+5.32\%}_{-4.63\%}$ & $106.94^{-0.06\%}_{-0.52\%}$ & 4.44\\
			\hline
	\end{tabular}
\caption{Cross-sections with scale uncertainties and the corresponding RIs due to NLO QCD corrections for the NC and CC processes for the benchmark values $(\lambda_{1Z},\lambda_{2Z},\widetilde{\lambda}_Z) = (0.4, 0.3, 0.5)$ and $(\lambda_{1W},\lambda_{2W},\widetilde{\lambda}_W) = (0.07, 0.02, 0.2)$.}
\label{table:bsm_results}
\end{center}
\end{table}

In Tab.~\ref{table:bsm_results}, we see that the cross-section for the NC process receives 2.9\% to 4.1\% QCD correction at NLO relative to the LO contribution for all three BSM parameters of $HZZ$ couplings. Similarly, the cross-section for the CC process receives around 4.4-4.7\% corrections for all three BSM parameters of $HWW$ couplings relative to the LO contribution. These corrections are consistent with the NLO corrections to the SM processes shown in Tab.~\ref{table:sm_results}.  At LO, the scale uncertainties are around  $\pm$ 5-6\% for both the NC and CC processes, similar to their associated SM predictions. At NLO, uncertainties go from nearly zero to  -(0.4-0.7)\% for both the NC and CC processes.

\begin{figure}[htp]
\begin{center}
\includegraphics [angle=0,width=0.3\linewidth]{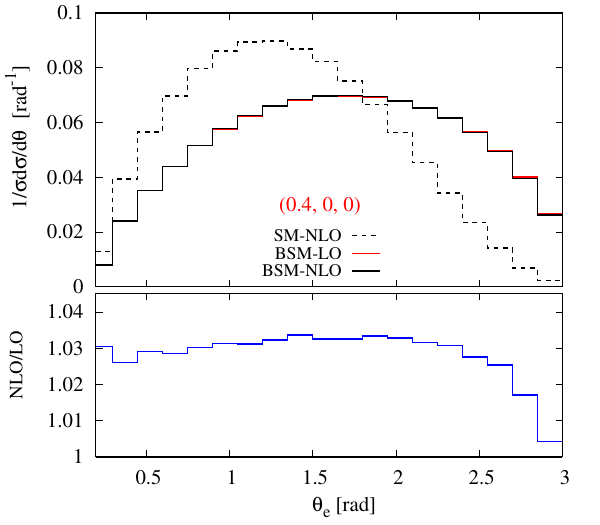}
\includegraphics [angle=0,width=0.3\linewidth]{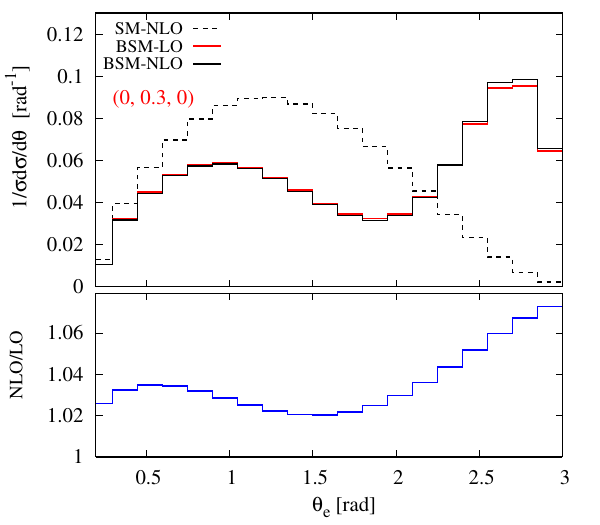}
\includegraphics [angle=0,width=0.3\linewidth]{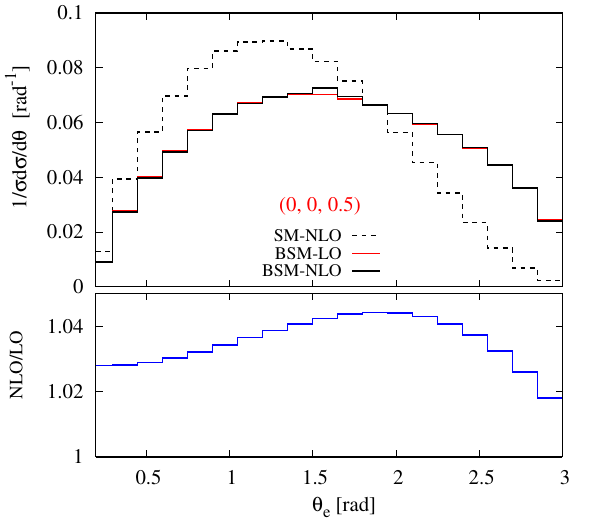}
\caption{ The LO (red) and NLO QCD (black) predictions for $\theta_e$ in the NC process for benchmark values $\lambda_{1Z} = 0.4$, $\lambda_{2Z} = 0.3$, and $\widetilde{\lambda}_Z = 0.5$ (left-to-right). For comparison, the SM results at NLO are also shown using dashed lines. The NLO K-factors (blue) are shown in the lower panels. }
\label{fig:nc_theta}
\end{center}
\end{figure}

\begin{figure}[htp]
\begin{center}
\includegraphics [angle=0,width=0.3\linewidth]{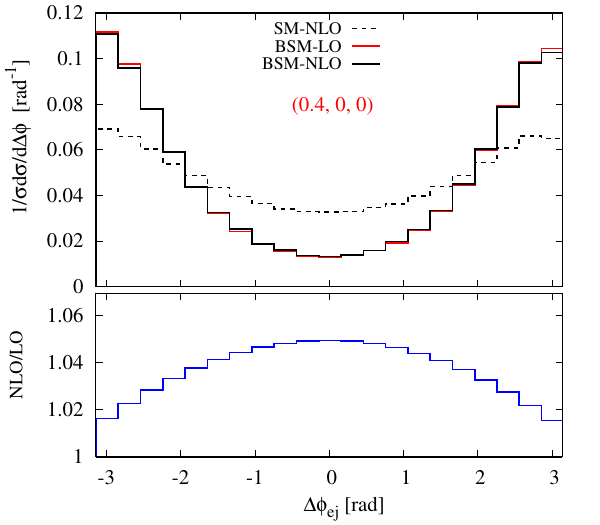}
\includegraphics [angle=0,width=0.3\linewidth]{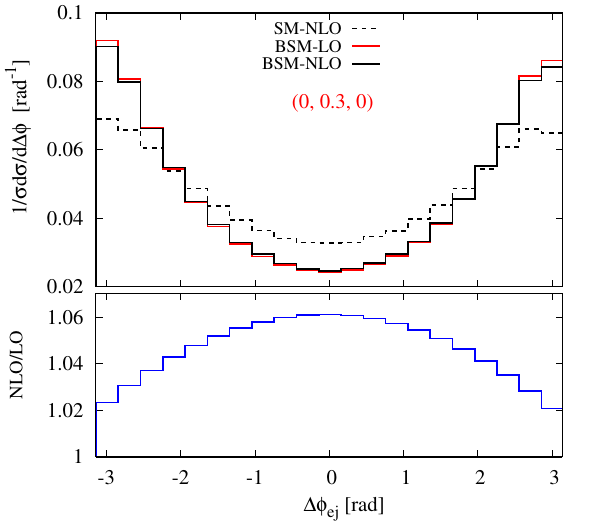}
\includegraphics [angle=0,width=0.3\linewidth]{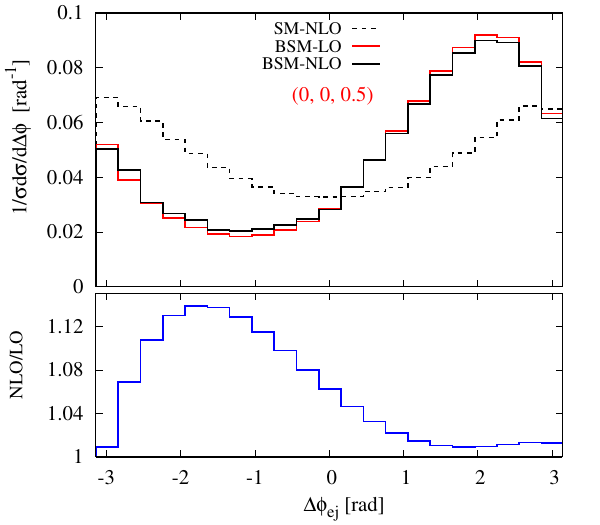}
\caption{ The LO (red) and NLO QCD (black) predictions for $\Delta \phi_{ej}$ in the NC process for benchmark values $\lambda_{1Z} = 0.4$, $\lambda_{2Z} = 0.3$ and $\widetilde{\lambda}_Z = 0.5$ (left-to-right). For comparison, the SM results at NLO are also shown using dashed lines. The NLO K-factors (blue) are shown in the lower panel.}
\label{fig:nc_dphi}
\end{center}
\end{figure}

As discussed in Ref.~\cite{Sharma:2025ceq}, the observables $\theta_e$ and $\Delta\phi_{ej}$ exhibit the highest sensitivity to non-standard $HZZ$ couplings, while $\Delta\phi_{\slashed{E}j}$ is most sensitive to deviations in the $HWW$ coupling. We, therefore, restrict our discussions in this section to these angular observables only. 
In our previous study, we found that $\theta_e$ is best suited for constraining the $\lambda_{2Z}$ parameter. 
As can be seen in Fig.~\ref{fig:nc_theta}, the presence of non-standard couplings shifts the $\theta_e$ distribution towards $\pi$ with respect to the SM prediction. 
For $\lambda_{1Z}$ and $\lambda_{2Z}$, the QCD corrections to $\theta_e$ range from 0.5\%-3.5\% and 2\%-7.5\%, respectively. In the case of $\widetilde{\lambda}_Z$, the corrections vary between 2\% and 4.5\% across the bins, with comparatively larger K-factor around central bins.

The $\Delta \phi_{ej}$ distribution being CP-sensitive observable has been found suitable for efficiently 
constraining the $\widetilde{\lambda}_Z$ parameter~\cite{Sharma:2025ceq}. In Fig.~\ref{fig:nc_dphi}, the QCD corrections in the presence of BSM parameters of the $HZZ$ coupling in the NC process are shown. For the $\lambda_{1Z}$ and $\lambda_{2Z}$, corrections are symmetric around the central bin. The corrections for $\lambda_{1Z}$ are 2\% in outer bins and go up to 5\% in the mid region. In the case of $\lambda_{2Z}$, the distribution has corrections as large as 6\% in central bins. For $\widetilde{\lambda}_Z$, the correction is about 1\% in the outer bins, while in certain bins corresponding to negative values of $\Delta \phi_{ej}$ it becomes significantly larger, reaching up to 14\%.

\begin{figure}[htp]
\begin{center}
\includegraphics [angle=0,width=0.3\linewidth]{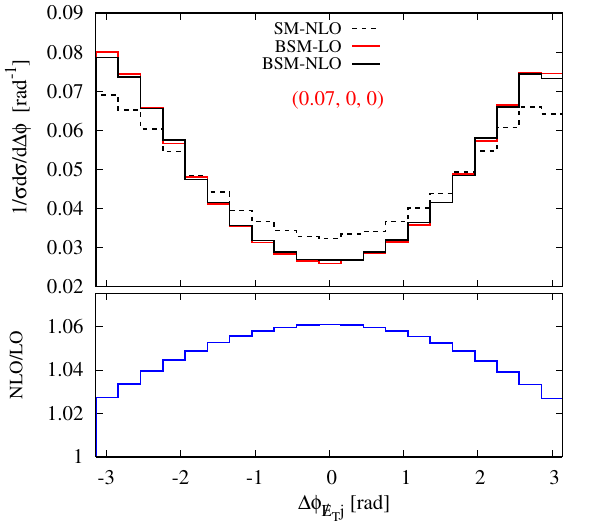}
\includegraphics [angle=0,width=0.3\linewidth]{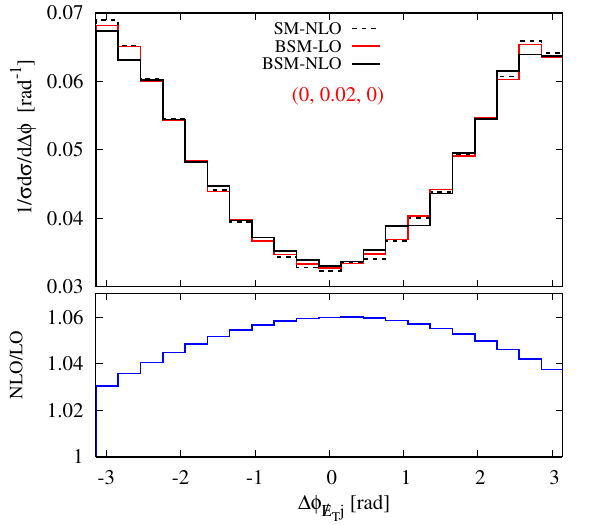}
\includegraphics [angle=0,width=0.3\linewidth]{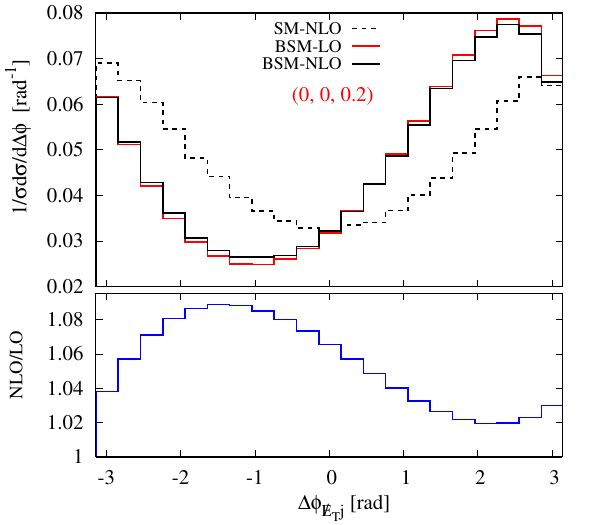}
\caption{ The LO (red) and NLO QCD (black) predictions for $\Delta \phi_{\slashed{E}j}$ in the CC process for benchmark values $\lambda_{1W} = 0.07$, $\lambda_{2W} = 0.02$ and $\widetilde{\lambda}_W = 0.2$ (left-to-right). For comparison, the SM results at NLO are also shown using dashed lines. The NLO K-factors (blue) are shown in the lower panel.}
\label{fig:cc_dphi}
\end{center}
\end{figure}

For the CC process, due to the non-observability of the final state neutrino, the only angular observable of interest is $\Delta \phi_{\slashed{E}j}$, which is most sensitive to $\widetilde{\lambda}_W$ parameter. The impact of QCD corrections on this observable is shown in Fig.~\ref{fig:cc_dphi} for the benchmark values of the non-standard $HWW$ interaction parameters. 
For each BSM parameter associated with a similar Lorentz structure of $HZZ$ and $HWW$ coupling, the pattern of correction in the $\Delta \phi_{\slashed{E}j}$ distribution is similar to that of the $\Delta \phi_{ej}$ distribution.  
We observe that the $\Delta \phi_{\slashed{E}j}$ distributions receive corrections of about 2–6\% for $\lambda_{1W}$ and $\lambda_{2W}$, while for $\widetilde{\lambda}_W$ the corrections range from 2–9\% across the bins.

\label{sub:np_results}

%%%%%%%%%%%%%%%%%%%%%%%%%%%%%%%%%%%%%%%%%%%%%%%%%%%%%%%%%%%%%%

\section{Conclusions}
\label{sec:concl}
In this work, we presented NLO QCD corrections to single Higgs production via the VBF process at  $ep$ collider with a 60~GeV electron beam and a 7~TeV proton beam. For the \texttt{NC} and \texttt{CC} processes with purely SM couplings, the NLO corrections enhance the total cross section by approximately 4.0\% and 4.6\%, respectively. We find that the corrections are PDF set dependent. A similar level of enhancement is observed when non-standard $HVV$ couplings are included. The scale uncertainty, which is about $\pm 5\%$ at LO, is significantly reduced to the permille level at NLO, which shows significant improvement in theoretical precision.
We have also studied the $K$-factor bin by bin for each distribution, which gives us more insight into the nature of QCD corrections in different phase-space regions. 

We also examined the impact of QCD corrections on key differential distributions relevant for event selection, such as the $p_T$, $\eta$, and invariant mass. Except for $\eta(j)$, all other kinematic observables receive modest corrections similar to the total cross section. The correction in the case of $\eta(j)$, however, can go beyond $\pm$20\% in the high rapidity regions.
Furthermore, we studied angular observables that are sensitive to non-standard $HVV$ interactions, including $\theta_e$, $\alpha_{ej}$, $\beta_{ejH}$ and $\Delta \phi_{ej}$ in the \texttt{NC} channel, and $\Delta \phi_{\slashed{E}j}$ in the \texttt{CC} channel. In both the standard model and in the presence of non-standard $HVV$ couplings, the corrections for these distributions are small, but the K-factors change non-trivially across the bins. 

Although the QCD corrections in the NC and CC are small, their precise impact on constraining the CP-even and CP-odd parameters of $HVV$ couplings requires a systematic inclusion of QCD corrections in background processes as well. Now that electroweak corrections are also available and they affect kinematic distributions significantly, their inclusion in the signal-background analysis is also desirable.

 \section*{Acknowledgements}
       BD would like to acknowledge Tousik Samui, Luca Buonocore, and Pankaj Agrawal for valuable discussions.
\bibliographystyle{JHEP}
\bibliography{ep2lhjref}

\providecommand{\href}[2]{#2}\begingroup\raggedright\begin{thebibliography}{10}

\bibitem{ATLAS:2012yve}
{\scshape ATLAS} collaboration, G.~Aad et~al., \emph{{Observation of a new
  particle in the search for the Standard Model Higgs boson with the ATLAS
  detector at the LHC}},
  \href{https://doi.org/10.1016/j.physletb.2012.08.020}{\emph{Phys. Lett. B}
  {\bfseries 716} (2012) 1} [\href{https://arxiv.org/abs/1207.7214}{{\ttfamily
  1207.7214}}].

\bibitem{CMS:2012qbp}
{\scshape CMS} collaboration, S.~Chatrchyan et~al., \emph{{Observation of a New
  Boson at a Mass of 125 GeV with the CMS Experiment at the LHC}},
  \href{https://doi.org/10.1016/j.physletb.2012.08.021}{\emph{Phys. Lett. B}
  {\bfseries 716} (2012) 30} [\href{https://arxiv.org/abs/1207.7235}{{\ttfamily
  1207.7235}}].

\bibitem{Higgs:1964pj}
P.~W. Higgs, \emph{{Broken Symmetries and the Masses of Gauge Bosons}},
  \href{https://doi.org/10.1103/PhysRevLett.13.508}{\emph{Phys. Rev. Lett.}
  {\bfseries 13} (1964) 508}.

\bibitem{Higgs:1964ia}
P.~W. Higgs, \emph{{Broken symmetries, massless particles and gauge fields}},
  \href{https://doi.org/10.1016/0031-9163(64)91136-9}{\emph{Phys. Lett.}
  {\bfseries 12} (1964) 132}.

\bibitem{Spira:2016ztx}
M.~Spira, \emph{{Higgs Boson Production and Decay at Hadron Colliders}},
  \href{https://doi.org/10.1016/j.ppnp.2017.04.001}{\emph{Prog. Part. Nucl.
  Phys.} {\bfseries 95} (2017) 98}
  [\href{https://arxiv.org/abs/1612.07651}{{\ttfamily 1612.07651}}].

\bibitem{LHCHiggsCrossSectionWorkingGroup:2016ypw}
{\scshape LHC Higgs Cross Section Working Group} collaboration, D.~de~Florian
  et~al., \emph{{Handbook of LHC Higgs Cross Sections: 4. Deciphering the
  Nature of the Higgs Sector}},
  \href{https://doi.org/10.23731/CYRM-2017-002}{\emph{CERN Yellow Rep. Monogr.}
  {\bfseries 2} (2017) 1} [\href{https://arxiv.org/abs/1610.07922}{{\ttfamily
  1610.07922}}].

\bibitem{CMS:2022dwd}
{\scshape CMS} collaboration, A.~Tumasyan et~al., \emph{{A portrait of the
  Higgs boson by the CMS experiment ten years after the discovery.}},
  \href{https://doi.org/10.1038/s41586-022-04892-x}{\emph{Nature} {\bfseries
  607} (2022) 60} [\href{https://arxiv.org/abs/2207.00043}{{\ttfamily
  2207.00043}}].

\bibitem{ATLAS:2022vkf}
{\scshape ATLAS} collaboration, G.~Aad et~al., \emph{{A detailed map of Higgs
  boson interactions by the ATLAS experiment ten years after the discovery}},
  \href{https://doi.org/10.1038/s41586-022-04893-w}{\emph{Nature} {\bfseries
  607} (2022) 52} [\href{https://arxiv.org/abs/2207.00092}{{\ttfamily
  2207.00092}}].

\bibitem{LHeC:2020van}
{\scshape LHeC, FCC-he Study Group} collaboration, P.~Agostini et~al.,
  \emph{{The Large Hadron\textendash{}Electron Collider at the HL-LHC}},
  \href{https://doi.org/10.1088/1361-6471/abf3ba}{\emph{J. Phys. G} {\bfseries
  48} (2021) 110501} [\href{https://arxiv.org/abs/2007.14491}{{\ttfamily
  2007.14491}}].

\bibitem{LHeCStudyGroup:2012zhm}
{\scshape LHeC Study Group} collaboration, J.~L. Abelleira~Fernandez et~al.,
  \emph{{A Large Hadron Electron Collider at CERN: Report on the Physics and
  Design Concepts for Machine and Detector}},
  \href{https://doi.org/10.1088/0954-3899/39/7/075001}{\emph{J. Phys. G}
  {\bfseries 39} (2012) 075001}
  [\href{https://arxiv.org/abs/1206.2913}{{\ttfamily 1206.2913}}].

\bibitem{Ahmadova:2025vzd}
F.~Ahmadova et~al., \emph{{The Large Hadron electron Collider as a bridge
  project for CERN}},  \href{https://arxiv.org/abs/2503.17727}{{\ttfamily
  2503.17727}}.

\bibitem{Stong:1993bq}
M.~L. Stong and K.~Hagiwara, \emph{{Probing the scalar sector in e+ e-
  ---\ensuremath{>} f anti-f phi}},  in \emph{{2nd International Workshop on
  Physics and Experiments with Linear e+ e- Colliders}}, 8, 1993,
  \href{https://arxiv.org/abs/hep-ph/9307255}{{\ttfamily hep-ph/9307255}}.

\bibitem{Dutta:2008bh}
S.~Dutta, K.~Hagiwara and Y.~Matsumoto, \emph{{Measuring the Higgs-Vector boson
  Couplings at Linear $e^{+} e^{-}$ Collider}},
  \href{https://doi.org/10.1103/PhysRevD.78.115016}{\emph{Phys. Rev. D}
  {\bfseries 78} (2008) 115016}
  [\href{https://arxiv.org/abs/0808.0477}{{\ttfamily 0808.0477}}].

\bibitem{Biswal:2012mp}
S.~S. Biswal, R.~M. Godbole, B.~Mellado and S.~Raychaudhuri, \emph{{Azimuthal
  Angle Probe of Anomalous $HWW$ Couplings at a High Energy $ep$ Collider}},
  \href{https://doi.org/10.1103/PhysRevLett.109.261801}{\emph{Phys. Rev. Lett.}
  {\bfseries 109} (2012) 261801}
  [\href{https://arxiv.org/abs/1203.6285}{{\ttfamily 1203.6285}}].

\bibitem{Alloul:2013naa}
A.~Alloul, B.~Fuks and V.~Sanz, \emph{{Phenomenology of the Higgs Effective
  Lagrangian via FEYNRULES}},
  \href{https://doi.org/10.1007/JHEP04(2014)110}{\emph{JHEP} {\bfseries 04}
  (2014) 110} [\href{https://arxiv.org/abs/1310.5150}{{\ttfamily 1310.5150}}].

\bibitem{Kumar:2015kca}
M.~Kumar, X.~Ruan, R.~Islam, A.~S. Cornell, M.~Klein, U.~Klein et~al.,
  \emph{{Probing anomalous couplings using di-Higgs production in
  electron\textendash{}proton collisions}},
  \href{https://doi.org/10.1016/j.physletb.2016.11.039}{\emph{Phys. Lett. B}
  {\bfseries 764} (2017) 247}
  [\href{https://arxiv.org/abs/1509.04016}{{\ttfamily 1509.04016}}].

\bibitem{Hernandez-Juarez:2023dor}
A.~I. Hern\'andez-Ju\'arez, G.~Tavares-Velasco and A.~Fern\'andez-T\'ellez,
  \emph{{New evaluation of the HZZ coupling: Direct bounds on anomalous
  contributions and CP-violating effects via a new asymmetry}},
  \href{https://doi.org/10.1103/PhysRevD.107.115031}{\emph{Phys. Rev. D}
  {\bfseries 107} (2023) 115031}
  [\href{https://arxiv.org/abs/2301.13127}{{\ttfamily 2301.13127}}].

\bibitem{Han:2009ra}
T.~Han and Y.~Li, \emph{{Genuine CP-odd Observables at the LHC}},
  \href{https://doi.org/10.1016/j.physletb.2009.12.047}{\emph{Phys. Lett. B}
  {\bfseries 683} (2010) 278}
  [\href{https://arxiv.org/abs/0911.2933}{{\ttfamily 0911.2933}}].

\bibitem{Christensen:2010pf}
N.~D. Christensen, T.~Han and Y.~Li, \emph{{Testing CP Violation in ZZH
  Interactions at the LHC}},
  \href{https://doi.org/10.1016/j.physletb.2010.08.008}{\emph{Phys. Lett. B}
  {\bfseries 693} (2010) 28} [\href{https://arxiv.org/abs/1005.5393}{{\ttfamily
  1005.5393}}].

\bibitem{Dwivedi:2016xwm}
S.~Dwivedi, D.~K. Ghosh, B.~Mukhopadhyaya and A.~Shivaji, \emph{{Distinguishing
  $CP$-odd couplings of the Higgs boson to weak boson pairs}},
  \href{https://doi.org/10.1103/PhysRevD.93.115039}{\emph{Phys. Rev. D}
  {\bfseries 93} (2016) 115039}
  [\href{https://arxiv.org/abs/1603.06195}{{\ttfamily 1603.06195}}].

\bibitem{PhysRevD.83.113004}
N.~Desai, B.~Mukhopadhyaya and D.~K. Ghosh, \emph{$cp$-violating $hww$
  couplings at the large hadron collider},
  \href{https://doi.org/10.1103/PhysRevD.83.113004}{\emph{Phys. Rev. D}
  {\bfseries 83} (2011) 113004}.

\bibitem{Plehn:2001nj}
T.~Plehn, D.~L. Rainwater and D.~Zeppenfeld, \emph{{Determining the Structure
  of Higgs Couplings at the LHC}},
  \href{https://doi.org/10.1103/PhysRevLett.88.051801}{\emph{Phys. Rev. Lett.}
  {\bfseries 88} (2002) 051801}
  [\href{https://arxiv.org/abs/hep-ph/0105325}{{\ttfamily hep-ph/0105325}}].

\bibitem{Hankele:2006ma}
V.~Hankele, G.~Klamke, D.~Zeppenfeld and T.~Figy, \emph{{Anomalous Higgs boson
  couplings in vector boson fusion at the CERN LHC}},
  \href{https://doi.org/10.1103/PhysRevD.74.095001}{\emph{Phys. Rev. D}
  {\bfseries 74} (2006) 095001}
  [\href{https://arxiv.org/abs/hep-ph/0609075}{{\ttfamily hep-ph/0609075}}].

\bibitem{Han:2000mi}
T.~Han and J.~Jiang, \emph{{CP violating Z Z H coupling at e+ e- linear
  colliders}}, \href{https://doi.org/10.1103/PhysRevD.63.096007}{\emph{Phys.
  Rev. D} {\bfseries 63} (2001) 096007}
  [\href{https://arxiv.org/abs/hep-ph/0011271}{{\ttfamily hep-ph/0011271}}].

\bibitem{Hagiwara:2000tk}
K.~Hagiwara, S.~Ishihara, J.~Kamoshita and B.~A. Kniehl, \emph{{Prospects of
  measuring general Higgs couplings at e+ e- linear colliders}},
  \href{https://doi.org/10.1007/s100520000366}{\emph{Eur. Phys. J. C}
  {\bfseries 14} (2000) 457}
  [\href{https://arxiv.org/abs/hep-ph/0002043}{{\ttfamily hep-ph/0002043}}].

\bibitem{Li:2019evl}
H.-D. Li, C.-D. L\"u and L.-Y. Shan, \emph{{Sensitivity study of anomalous
  $HZZ$ couplings at a future Higgs factory}},
  \href{https://doi.org/10.1088/1674-1137/43/10/103001}{\emph{Chin. Phys. C}
  {\bfseries 43} (2019) 103001}
  [\href{https://arxiv.org/abs/1901.10218}{{\ttfamily 1901.10218}}].

\bibitem{Hernandez-Juarez:2024zpk}
A.~I. Hern{\'a}ndez-Ju{\'a}rez, R.~Gait{\'a}n and G.~Tavares-Velasco,
  \emph{{Polarized and unpolarized off-shell ${ \boldsymbol H^\ast{\bf\to}
  \boldsymbol Z\boldsymbol Z{\bf\rightarrow 4}\boldsymbol\ell}$ decay above the
  ${{\bf 2}\boldsymbol m_Z} $ threshold*}},
  \href{https://doi.org/10.1088/1674-1137/ad62de}{\emph{Chin. Phys. C}
  {\bfseries 48} (2024) 113103}
  [\href{https://arxiv.org/abs/2402.18497}{{\ttfamily 2402.18497}}].

\bibitem{Hernandez-Juarez:2025nzh}
A.~I. Hern{\'a}ndez-Ju{\'a}rez and R.~Gait{\'a}n, \emph{{CP violation in the
  HZZ vertex and left-right asymmetries}},
  \href{https://doi.org/10.1103/g124-ttp5}{\emph{Phys. Rev. D} {\bfseries 112}
  (2025) 033006} [\href{https://arxiv.org/abs/2501.18807}{{\ttfamily
  2501.18807}}].

\bibitem{Sharma:2022epc}
P.~Sharma and A.~Shivaji, \emph{{Probing non-standard HVV (V = W, Z) couplings
  in single Higgs production at future electron-proton collider}},
  \href{https://doi.org/10.1007/JHEP10(2022)108}{\emph{JHEP} {\bfseries 10}
  (2022) 108} [\href{https://arxiv.org/abs/2207.03862}{{\ttfamily
  2207.03862}}].

\bibitem{Sharma:2025ceq}
P.~Sharma and A.~Shivaji, \emph{{Role of angular observables in probing
  non-standard $HZZ$ couplings at an electron-proton collider}},
  \href{https://arxiv.org/abs/2506.02558}{{\ttfamily 2506.02558}}.

\bibitem{Sharma:2024ujn}
P.~Sharma, B.~Das and A.~Shivaji, \emph{{NLO QCD effects on angular observables
  in single Higgs production at electron-proton collider}},
  \href{https://doi.org/10.22323/1.476.0085}{\emph{PoS} {\bfseries ICHEP2024}
  (2025) 085} [\href{https://arxiv.org/abs/2411.07950}{{\ttfamily
  2411.07950}}].

\bibitem{Blumlein:1992eh}
J.~Blumlein, G.~J. van Oldenborgh and R.~Ruckl, \emph{{QCD and QED corrections
  to Higgs boson production in charged current e p scattering}},
  \href{https://doi.org/10.1016/0550-3213(93)90207-6}{\emph{Nucl. Phys. B}
  {\bfseries 395} (1993) 35}
  [\href{https://arxiv.org/abs/hep-ph/9209219}{{\ttfamily hep-ph/9209219}}].

\bibitem{Jager:2010zm}
B.~Jager, \emph{{Next-to-leading order QCD corrections to Higgs production at a
  future lepton-proton collider}},
  \href{https://doi.org/10.1103/PhysRevD.81.054018}{\emph{Phys. Rev. D}
  {\bfseries 81} (2010) 054018}
  [\href{https://arxiv.org/abs/1001.3789}{{\ttfamily 1001.3789}}].

\bibitem{Wang:2022awk}
B.~Wang, K.~Wang and H.~Xiong, \emph{{Electroweak corrections to Higgs boson
  production via W W fusion at the future LHeC}},
  \href{https://doi.org/10.1103/PhysRevD.106.093011}{\emph{Phys. Rev. D}
  {\bfseries 106} (2022) 093011}
  [\href{https://arxiv.org/abs/2207.14515}{{\ttfamily 2207.14515}}].

\bibitem{Xiong:2023qnr}
H.~Xiong, H.~Hou, Z.~Qian, Q.~Xu and B.~Wang, \emph{{Electroweak corrections to
  Higgs boson production via Z Z fusion at the future LHeC}},
  \href{https://doi.org/10.1103/PhysRevD.109.116021}{\emph{Phys. Rev. D}
  {\bfseries 109} (2024) 116021}
  [\href{https://arxiv.org/abs/2311.14912}{{\ttfamily 2311.14912}}].

\bibitem{Sherpa:2024mfk}
{\scshape Sherpa} collaboration, E.~Bothmann et~al., \emph{{Event generation
  with Sherpa 3}}, \href{https://doi.org/10.1007/JHEP12(2024)156}{\emph{JHEP}
  {\bfseries 12} (2024) 156}
  [\href{https://arxiv.org/abs/2410.22148}{{\ttfamily 2410.22148}}].

\bibitem{Alwall:2014hca}
J.~Alwall, R.~Frederix, S.~Frixione, V.~Hirschi, F.~Maltoni, O.~Mattelaer
  et~al., \emph{{The automated computation of tree-level and next-to-leading
  order differential cross sections, and their matching to parton shower
  simulations}}, \href{https://doi.org/10.1007/JHEP07(2014)079}{\emph{JHEP}
  {\bfseries 07} (2014) 079} [\href{https://arxiv.org/abs/1405.0301}{{\ttfamily
  1405.0301}}].

\bibitem{Peskin:2011in}
M.~E. Peskin, \emph{{Simplifying Multi-Jet QCD Computation}},  in \emph{{13th
  Mexican School of Particles and Fields}}, 1, 2011,
  \href{https://arxiv.org/abs/1101.2414}{{\ttfamily 1101.2414}}.

\bibitem{KLEISS1986159}
R.~Kleiss and W.~Stirling, \emph{Cross sections for the production of an
  arbitrary number of photons in electron-positron annihilation},
  \href{https://doi.org/https://doi.org/10.1016/0370-2693(86)90454-5}{\emph{Physics
  Letters B} {\bfseries 179} (1986) 159}.

\bibitem{Agrawal:2021owa}
P.~Agrawal and B.~Das, \emph{{W+W\ensuremath{-}H production through bottom
  quarks fusion at hadron colliders}},
  \href{https://doi.org/10.1016/j.physletb.2021.136461}{\emph{Phys. Lett. B}
  {\bfseries 820} (2021) 136461}
  [\href{https://arxiv.org/abs/2103.02682}{{\ttfamily 2103.02682}}].

\bibitem{Das:2023gsh}
B.~Das, \emph{{Probing Anomalous Higgs Boson Couplings at Colliders}}, Ph.D.
  thesis, Bhubaneswar, Inst. Phys., 2023.
\newblock \href{https://arxiv.org/abs/2409.01264}{{\ttfamily 2409.01264}}.

\bibitem{THOOFT1972189}
G.~{'t Hooft} and M.~Veltman, \emph{Regularization and renormalization of gauge
  fields},
  \href{https://doi.org/https://doi.org/10.1016/0550-3213(72)90279-9}{\emph{Nuclear
  Physics B} {\bfseries 44} (1972) 189}.

\bibitem{Kleiss:1985gy}
R.~Kleiss, W.~J. Stirling and S.~D. Ellis, \emph{{A New Monte Carlo Treatment
  of Multiparticle Phase Space at High-energies}},
  \href{https://doi.org/10.1016/0010-4655(86)90119-0}{\emph{Comput. Phys.
  Commun.} {\bfseries 40} (1986) 359}.

\bibitem{Veseli:1997hr}
S.~Veseli, \emph{{Multidimensional integration in a heterogeneous network
  environment}},
  \href{https://doi.org/10.1016/S0010-4655(97)00120-3}{\emph{Comput. Phys.
  Commun.} {\bfseries 108} (1998) 9}
  [\href{https://arxiv.org/abs/physics/9710017}{{\ttfamily physics/9710017}}].

\bibitem{Lepage:1977sw}
G.~Lepage, \emph{{A New Algorithm for Adaptive Multidimensional Integration}},
  \href{https://doi.org/10.1016/0021-9991(78)90004-9}{\emph{J. Comput. Phys.}
  {\bfseries 27} (1978) 192}.

\bibitem{10.7551/mitpress/5712.001.0001}
A.~Geist, A.~Beguelin, J.~Dongarra, W.~Jiang, R.~Manchek and V.~S. Sunderam,
  \emph{PVM: A Users' Guide and Tutorial for Network Parallel Computing}. The
  MIT Press, 11, 1994,
  \href{https://doi.org/10.7551/mitpress/5712.001.0001}{10.7551/mitpress/5712.001.0001}.

\bibitem{Dittmaier:2003bc}
S.~Dittmaier, \emph{{Separation of soft and collinear singularities from one
  loop N point integrals}},
  \href{https://doi.org/10.1016/j.nuclphysb.2003.10.003}{\emph{Nucl. Phys. B}
  {\bfseries 675} (2003) 447}
  [\href{https://arxiv.org/abs/hep-ph/0308246}{{\ttfamily hep-ph/0308246}}].

\bibitem{Catani:1996vz}
S.~Catani and M.~H. Seymour, \emph{{A General algorithm for calculating jet
  cross-sections in NLO QCD}},
  \href{https://doi.org/10.1016/S0550-3213(96)00589-5}{\emph{Nucl. Phys. B}
  {\bfseries 485} (1997) 291}
  [\href{https://arxiv.org/abs/hep-ph/9605323}{{\ttfamily hep-ph/9605323}}].

\bibitem{Cacciari:2011ma}
M.~Cacciari, G.~P. Salam and G.~Soyez, \emph{{FastJet User Manual}},
  \href{https://doi.org/10.1140/epjc/s10052-012-1896-2}{\emph{Eur. Phys. J. C}
  {\bfseries 72} (2012) 1896}
  [\href{https://arxiv.org/abs/1111.6097}{{\ttfamily 1111.6097}}].

\bibitem{Cacciari:2005hq}
M.~Cacciari and G.~P. Salam, \emph{{Dispelling the $N^{3}$ myth for the $k_t$
  jet-finder}},
  \href{https://doi.org/10.1016/j.physletb.2006.08.037}{\emph{Phys. Lett. B}
  {\bfseries 641} (2006) 57}
  [\href{https://arxiv.org/abs/hep-ph/0512210}{{\ttfamily hep-ph/0512210}}].

\bibitem{ParticleDataGroup:2024cfk}
{\scshape Particle Data Group} collaboration, S.~Navas et~al., \emph{{Review of
  particle physics}},
  \href{https://doi.org/10.1103/PhysRevD.110.030001}{\emph{Phys. Rev. D}
  {\bfseries 110} (2024) 030001}.

\bibitem{Buckley:2014ana}
A.~Buckley, J.~Ferrando, S.~Lloyd, K.~Nordstr\"om, B.~Page, M.~R\"ufenacht
  et~al., \emph{{LHAPDF6: parton density access in the LHC precision era}},
  \href{https://doi.org/10.1140/epjc/s10052-015-3318-8}{\emph{Eur. Phys. J. C}
  {\bfseries 75} (2015) 132} [\href{https://arxiv.org/abs/1412.7420}{{\ttfamily
  1412.7420}}].

\bibitem{Guzzi:2022rca}
M.~Guzzi, T.~J. Hobbs, K.~Xie, J.~Huston, P.~Nadolsky and C.~P. Yuan,
  \emph{{The persistent nonperturbative charm enigma}},
  \href{https://doi.org/10.1016/j.physletb.2023.137975}{\emph{Phys. Lett. B}
  {\bfseries 843} (2023) 137975}
  [\href{https://arxiv.org/abs/2211.01387}{{\ttfamily 2211.01387}}].

\bibitem{Hou:2019efy}
T.-J. Hou et~al., \emph{{New CTEQ global analysis of quantum chromodynamics
  with high-precision data from the LHC}},
  \href{https://doi.org/10.1103/PhysRevD.103.014013}{\emph{Phys. Rev. D}
  {\bfseries 103} (2021) 014013}
  [\href{https://arxiv.org/abs/1912.10053}{{\ttfamily 1912.10053}}].

\bibitem{NNPDF:2021njg}
{\scshape NNPDF} collaboration, R.~D. Ball et~al., \emph{{The path to proton
  structure at 1\% accuracy}},
  \href{https://doi.org/10.1140/epjc/s10052-022-10328-7}{\emph{Eur. Phys. J. C}
  {\bfseries 82} (2022) 428}
  [\href{https://arxiv.org/abs/2109.02653}{{\ttfamily 2109.02653}}].

\end{thebibliography}\endgroup

\end{document}